\documentclass[a4paper]{article}
\usepackage[utf8]{inputenc}
\usepackage[T1]{fontenc}
\usepackage{hyphenat}
\usepackage{xspace}
\usepackage{amsmath}
\usepackage{amsfonts}
\usepackage{hyperref}
\usepackage{url}
\usepackage{booktabs}
\usepackage{multirow}
\usepackage{makecell}
\usepackage{caption}
\usepackage{minibox}
\usepackage{bbm}
\usepackage{graphicx}
\usepackage{balance}
\usepackage{mathtools}
\usepackage{color}
\usepackage{marvosym}
\usepackage{ifthen}
\usepackage{textcomp}
\usepackage{enumitem}
\usepackage{verbatim}
\usepackage{algorithm}
\usepackage{numprint}
\usepackage{balance}

\usepackage{amsthm}
\theoremstyle{plain}

\newcommand{\chatoDisplayMode}[1]{#1}

\definecolor{MyRed}{rgb}{0.6,0.0,0.0} 
\definecolor{MyBlack}{rgb}{0.1,0.1,0.1} 
\newcommand{\inred}[1]{{\color{MyRed}\sf\textbf{\textsc{#1}}}}
\newcommand{\frameit}[2]{
  \begin{center}
  {\color{MyRed}
  \framebox[.9\columnwidth][l]{
    \begin{minipage}{.85\columnwidth}
    \inred{#1}: {\sf\color{MyBlack}#2}
    \end{minipage}
  }\\
  }
  \end{center}
}

\newcommand{\note}[2][]{\chatoDisplayMode{\def\@tmpsig{#1}\frameit{{\Pointinghand} Note}{#2\ifx \@tmpsig \@empty \else \mbox{ --\em #1}\fi}}}
\newcommand{\todo}[2][]{\chatoDisplayMode{\def\@tmpsig{#1}\frameit{{\Writinghand} To-do}{#2\ifx \@tmpsig \@empty \else \mbox{ --\em #1}\fi}}}

\newcommand{\abbrevStyle}[1]{#1}
\newcommand{\ie}{\abbrevStyle{i.e.}\xspace}
\newcommand{\eg}{\abbrevStyle{e.g.}\xspace}
\newcommand{\cf}{\abbrevStyle{cf.}\xspace}

\newcommand{\vs}{\abbrevStyle{vs.}\xspace}

\newcommand{\Secref}[1]{Sec.~\ref{#1}}

\newcommand{\Tabref}[1]{Table~\ref{#1}}
\newcommand{\Figref}[1]{Fig.~\ref{#1}}

\newcommand{\xhdr}[1]{\vspace{1.7mm}\noindent{{\bf #1.}}}

\newcommand{\textcite}[1]{\citeauthor{#1} \shortcite{#1}}

\newcommand{\hide}[1]{}

\makeatletter
\newcommand{\iffont}[2]{\ifthenelse{\equal{\f@family}{#1}}{#2}{}}
\makeatother

\iffont{ptm}{
  \usepackage{mathptmx}

  \DeclareSymbolFont{greek}{OML}{cmm}{m}{n}
  \DeclareMathSymbol{\alpha}{\mathalpha}{greek}{"0B}
  \DeclareMathSymbol{\beta}{\mathalpha}{greek}{"0C}
  \DeclareMathSymbol{\gamma}{\mathalpha}{greek}{"0D}
  \DeclareMathSymbol{\delta}{\mathalpha}{greek}{"0E}
  \DeclareMathSymbol{\epsilon}{\mathalpha}{greek}{"0F}
  \DeclareMathSymbol{\zeta}{\mathalpha}{greek}{"10}
  \DeclareMathSymbol{\eta}{\mathalpha}{greek}{"11}
  \DeclareMathSymbol{\theta}{\mathalpha}{greek}{"12}
  \DeclareMathSymbol{\iota}{\mathalpha}{greek}{"13}
  \DeclareMathSymbol{\kappa}{\mathalpha}{greek}{"14}
  \DeclareMathSymbol{\lambda}{\mathalpha}{greek}{"15}
  \DeclareMathSymbol{\mu}{\mathalpha}{greek}{"16}
  \DeclareMathSymbol{\nu}{\mathalpha}{greek}{"17}
  \DeclareMathSymbol{\xi}{\mathalpha}{greek}{"18}
  \DeclareMathSymbol{\pi}{\mathalpha}{greek}{"19}
  \DeclareMathSymbol{\rho}{\mathalpha}{greek}{"1A}
  \DeclareMathSymbol{\sigma}{\mathalpha}{greek}{"1B}
  \DeclareMathSymbol{\tau}{\mathalpha}{greek}{"1C}
  \DeclareMathSymbol{\upsilon}{\mathalpha}{greek}{"1D}
  \DeclareMathSymbol{\phi}{\mathalpha}{greek}{"1E}
  \DeclareMathSymbol{\chi}{\mathalpha}{greek}{"1F}
  \DeclareMathSymbol{\psi}{\mathalpha}{greek}{"20}
  \DeclareMathSymbol{\omega}{\mathalpha}{greek}{"21}
  \DeclareMathSymbol{\varepsilon}{\mathalpha}{greek}{"22}
  \DeclareMathSymbol{\vartheta}{\mathalpha}{greek}{"23}
  \DeclareMathSymbol{\varpi}{\mathalpha}{greek}{"24}
  \DeclareMathSymbol{\varrho}{\mathalpha}{greek}{"25}
  \DeclareMathSymbol{\varsigma}{\mathalpha}{greek}{"26}
  \DeclareMathSymbol{\varphi}{\mathalpha}{greek}{"27}
  \DeclareSymbolFont{otone}{OT1}{cmr}{m}{n}
  \DeclareMathSymbol{\Gamma}{\mathalpha}{otone}{0}
  \DeclareMathSymbol{\Delta}{\mathalpha}{otone}{1}
  \DeclareMathSymbol{\Theta}{\mathalpha}{otone}{2}
  \DeclareMathSymbol{\Lambda}{\mathalpha}{otone}{3}
  \DeclareMathSymbol{\Xi}{\mathalpha}{otone}{4}
  \DeclareMathSymbol{\Pi}{\mathalpha}{otone}{5}
  \DeclareMathSymbol{\Sigma}{\mathalpha}{otone}{6}
  \DeclareMathSymbol{\Upsilon}{\mathalpha}{otone}{7}
  \DeclareMathSymbol{\Phi}{\mathalpha}{otone}{8}
  \DeclareMathSymbol{\Psi}{\mathalpha}{otone}{9}
  \DeclareMathSymbol{\Omega}{\mathalpha}{otone}{10}
  \DeclareSymbolFont{syms}{OML}{cmm}{m}{it}
  \DeclareMathSymbol{\partial}{\mathord}{syms}{"40}
  \DeclareMathAlphabet{\mathbold}{OML}{cmm}{b}{it}
  \DeclareSymbolFont{largesymbols}{OMX}{cmex}{m}{n}

}

\usepackage{tikz}
\usetikzlibrary{positioning, calc, shapes.geometric, shapes, shapes.multipart, arrows.meta, arrows, decorations.markings, external, trees}
\tikzset{
  node distance=0.5cm, % specifies the minimum distance between two nodes. Change if necessary.
}
\tikzstyle{Arrow} = [
  thick, 
  decoration={
    markings,
    mark=at position 1 with {
      \arrow[thick]{latex}
    }
  }, 
  shorten >= 0.5pt, preaction = {decorate},
  ]
\usepackage{amsmath}
\usepackage{amsthm}
\usepackage{amssymb}

\usepackage{xcolor}
\usepackage{silence}
\usepackage[pages=all, color=black, position={current page.south}, placement=bottom, scale=1, opacity=1, vshift=5mm]{background}

\newcommand{\ApplyGradient}[1]{%
        \ifdim #1 pt > \MidNumber pt
            \pgfmathsetmacro{\PercentColor}{max(min(100.0*(#1 - \MidNumber)/(\MaxNumber-\MidNumber),100.0),0.00)} %
            \hspace{-0.33em}\colorbox{green!\PercentColor!yellow}{#1}
        \else
            \pgfmathsetmacro{\PercentColor}{max(min(100.0*(\MidNumber - #1)/(\MidNumber-\MinNumber),100.0),0.00)} %
            \hspace{-0.33em}\colorbox{red!\PercentColor!yellow}{#1}
        \fi
}

\usepackage[margin=1in]{geometry} % full-width
\usepackage{subcaption}

\usepackage[utf8]{inputenc}
\usepackage{hyperref}
\hypersetup{
	unicode,
	pdfauthor={Author One, Author Two, Author Three},
	pdftitle={A simple article template},
	pdfsubject={A simple article template},
	pdfkeywords={article, template, simple},
	pdfproducer={LaTeX},
	pdfcreator={pdflatex}
}

\usepackage[sort&compress,numbers,square]{natbib}
\usepackage{amsmath}               
  {
      \theoremstyle{plain}
      \newtheorem{assumption}{Assumption}
  }
  
\usepackage{graphicx, color}
\graphicspath{{fig/}}

\usepackage{algorithm, algpseudocode} % use algorithm and algorithmicx for typesetting algorithms
\usepackage{mathrsfs} % for \mathscr command
\usepackage{lipsum}

\title{Food Choice Mimicry on a Large University Campus}

\author{Kristina Gligori\'{c} PhD$^1$\thanks{Corresponding author. Work done while at EPFL.},
 Arnaud Chiolero MD PhD$^3$\thanks{Also affiliated with Institute of Primary Health Care (BIHAM), University of Bern, Switzerland, and School of Population and Global Health, McGill University, Canada.} ,
 Emre K{\i}c{\i}man PhD$^4$, \\
 Ryen W. White PhD$^4$,
 Eric Horvitz MD PhD$^4$,
Robert West PhD$^2$\thanks{Corresponding author.}}

\date{$^1$Computer Science Department, Stanford University, Stanford, CA, United States \\ \texttt{gligoric@stanford.edu}\\%
	$^2$School of Computer and Communication Sciences, EPFL, Switzerland \\ \texttt{robert.west@epfl.ch}\\%
	$^3$Population Health Laboratory, University of Fribourg,  Switzerland  \texttt{arnaud.chiolero@unifr.ch}\\%
	$^4$Microsoft Research, Redmond, WA, United States\\
	\texttt{\{emrek,ryenw,horvitz\}@microsoft.com}\\[2ex]%
}

\begin{document}
\maketitle

\begin{abstract}
\noindent
Social influence is a strong determinant of food consumption, which in turn influences health. Although consistent observations have been made on the role of social factors in driving similarities in food consumption, much less is known about the precise governing mechanisms. We study social influence on food choice through carefully designed causal analyses, leveraging the sequential nature of shop queues on a major university campus. In particular, we consider a large number of adjacent purchases where a focal user immediately follows another user (``partner'') in the checkout queue and both make a purchase. Identifying the partner's impact on the focal user, we find strong evidence of a specific behavioral mechanism for how dietary similarities between individuals arise: {\em purchasing mimicry}, a phenomenon where the focal user copies the partner's purchases. For instance, across food additions purchased during lunchtime together with a meal, we find that the focal user is significantly more likely to purchase the food item when the partner buys the item, \vs when the partner does not, increasing the purchasing probability by {14\% in absolute terms}, or by {83\% in relative terms}. The effect is observed across all food types, but largest for condiments, and smallest for soft drinks. We find that no such effect is observed when a focal user is compared to a random (rather than directly preceding) partner. Furthermore, purchasing mimicry is present across age, gender, and status subpopulations, but strongest for students and the youngest persons. Finally, we find a dose--response relationship whereby mimicry decreases as proximity in the purchasing queue decreases. The results of this study elucidate the behavioral mechanism of purchasing mimicry and have further implications for understanding and improving dietary behaviors on campus. 

\end{abstract}

\section{Introduction}

%What is the problem and why is it interesting and important?

Diet is a critical factor in health \cite{delaney2011food,gakidou2017global,gittelsohn2012interventions}. As a consequence, behavioral interventions \cite{zhao2022computational} and policies that promote healthier diets are a public-health priority \cite{willett2019food}. Since social influence is known to be a strong determinant of food consumption \cite{collins2019two,mollen2013healthy}, research has explored the potential of social norms for designing public health interventions to change diets \cite{robinson_blissett_higgs_2013,ROBINSON2014414}, \eg, by promoting healthy dietary habits and physical activity \cite{aral2017exercise,fjeldsoe2011systematic}, losing weight \cite{jeffery1993strengthening,EISENBERG20051165}, and reducing food waste~\cite{reynolds2019consumption}.

In university campus environments in particular, students and staff consume meals regularly and in large quantities, impacting health and the environment. Universities therefore provide an opportunity to study food choice, with implications for the general population. Food consumption on campus is particularly consequential since university education coincides with adolescents' and young adults' transition into adulthood. During this period, new dietary habits can be formed, and it is a critical period to stay on a healthy track to reduce the risk of chronic diseases, such as obesity, diabetes, cardiovascular diseases, and cancer \cite{singh2008tracking,reilly2011long}. Since campuses are training and working environments, university food consumption is also an occupational health issue. Therefore, it is necessary to understand factors influencing behaviors in these environments, and understanding factors can, in turn, inform interventions and policies among university students and staff \cite{roy2019exploring}.

%Existing work, what is known so far? Current evidence?

A large body of prior work has consistently observed similarities between connected persons in social networks \cite{gligoric2021ties}, \eg, friends \cite{fletcher2011you} and family \cite{eurpub,pedersen2015following}, in a number of experimental and survey‐based studies~\cite{madan2010social,finnerty2010effects,patrick2005review,salvy2008effects,stevenson2007adolescents,harmon2016perceived}. The food choices of others have been observed to influence food choices~\cite{christie2018vegetarian,robinson2013food,hetherington2006situational,higgs2016social,shepherd_1999,levy2021social}, through perceived eating norms (\ie, behavioral similarity guided by beliefs about how others around us behave), and through modeling of food choices and intake (\ie, behavioral similarity guided by being exposed to how others around us behave)~\cite{robinson_blissett_higgs_2013}. Particular focus has been placed on unhealthy behaviors and their social influences \cite{CRUWYS20153,blok2013unhealthy}, observing that obesity \cite{christakis2007spread}, overeating \cite{doi:10.1086/644611}, fast food \cite{thornton2013barriers}, high-fat food \cite{FEUNEKES1998645,hermans2009effects}, and alcohol and snack consumption \cite{pachucki2011social,wouters2010peer} are impacted by social norms.

%What is the knowledge gap?

However, although similarities in food consumption driven by social factors have been consistently observed, much less is known about the precise governing mechanisms. There are numerous mechanisms postulated about how others influence our food consumption, including the processes of information gathering, minimizing regret, and integration concerns~\cite{robinson_blissett_higgs_2013}. Such mechanisms can result in both dish variety seeking and dish uniformity seeking~\cite{ariely2000sequential,de2013adolescents,munt2017barriers}. 

%too specific 
%Uniformity seeking was tested across a range of studies, for instance, in the form of uniformity by matching the other's food choice \cite{cavazza2017portion}. Conversely, variety seeking \cite{sharma2010impulse,ratner2002impact} has also been observed, since individuals make choices that diverge from those of others, to effectively communicate the desired identity~\cite{berger2007consumers}.

One potentially important mechanism of interpersonal influence on eating behavior may be \emph{behavioral mimicry}, referring to when a person copies the behavior of another. For instance, individuals automatically mimic the gestures and hand movements of others, as an unconscious attempt to make the other individual like them, since mimicry eases social interactions \cite{chartrand1999chameleon,chartrand2009human}. It has been shown that viewing another individual performing an action activates an immediate reaction in an individual's motor system \cite{iacoboni1999cortical,hermans2012mimicry}. Behavioral mimicry is also consistent with the least-effort principle when making choices---sometimes, simply doing what others do is the easiest choice \cite{zipf2016human}. Since eating is often habitual, \ie, automatically driven by external cues, unconscious behavioral mimicry may be a key interpersonal influence mechanism when eating with others. 

% What we ask specifically and why is answering these questions important?

Previous studies have shown evidence of mimicry in behaviors linked to food consumption: people tend to adjust their intake directly to their eating companions by eating more when others eat more and less when others eat less \cite{chartrand2013antecedents,tanner2008chameleons}. However, several questions remain unanswered: How prominent is food purchasing mimicry? What foods are the most associated with purchasing mimicry, and what subpopulations are the most affected? Identifying the role of purchasing mimicry in social norms is the first necessary step toward determining whether and to which extent purchasing mimicry can be leveraged for behavioral interventions.

%Why is it hard to fill it? (E.g., how do previous approaches fail?), why hasn't it been solved before? (Or, what's wrong with previously proposed solutions?)

Despite the postulated importance of social factors, identifying and measuring mimicry in food consumption remains challenging. On the one hand, experimental studies monitor behaviors in artificial settings where people are aware they are being observed \cite{sullivan2021healthful}, which involves participation effect challenges, referred to as the Hawthorne effect \cite{mccambridge2014systematic,tiefenbeck2016magnitude}. Furthermore, experimental studies to date have been limited to observing people in small\hyp scale scenarios with a short duration, often in a laboratory setting \cite{ROBINSON2014414,robinson2013food}.
Most notably, such studies rely on confederate design, testing whether pairing a participant with an actor (the ``confederate'') influences the amount and type \cite{robinson2013food,bell2019sensing} of food eaten by the participant and their biting pattern \cite{sharps2015examining}, \ie, whether individuals take a bite of their meal in congruence with their eating companion rather than eating at their own pace \cite{hermans2012mimicry}. More naturalistic experimental settings attempt to increase the validity of the findings by instructing participants to perform an unrelated activity while food is provided and consumption patterns are recorded \cite{hermans2009effects}.

On the other hand, observational studies face limitations due to the presence of confounding factors and biases. In real\hyp world settings, it remains challenging to measure and disentangle properties that are relevant in the context of food consumption, such as attributes of the individuals and the environment. Another challenge is homophily, people's tendency to form ties with others similar to themselves to begin with \cite{aral2009distinguishing,shalizi2011homophily,shalizi2016estimating,kossinets2006empirical,de2011homophily,levy2021social}.

At present, characterizations of food purchasing mimicry originate from different experimental conditions. More fine-grained data sources and design paradigms are needed to identify behavioral mimicry and how it varies across foods and subpopulations, since aggregate insights may not reflect the true effect equally well for everyone~\cite{althoff2022large}. Researchers have only recently been addressing gaps in the knowledge about human dietary behaviors by studying digital traces in the context of food consumption to measure factors of well-being related to nutrition \cite{gligoric2021ties} and beyond~\cite{barclay2013peer,madan2010social,nook2015social,sefidgar2019passively,swain2020leveraging,gligoric2022biased,gligoric2022population}.

%Our work, how does our study differ
\xhdr{Our approach} The present study addresses the challenges of understanding the role of mimicry, in the case of a university campus environment. We leverage a large\hyp data set of shop records that captures the order of food selection and purchasing, allowing us to measure whether early customers influence late customers. In particular, we analyze an anonymized dataset of food purchases made on the EPFL university campus. The data spans from 2010 to 2018 and contains 18 million purchases made with a badge that allows anonymous linking to a person's purchase history and basic demographics. 

\begin{figure}[t!]
\centering
    \includegraphics[width=0.9\textwidth]{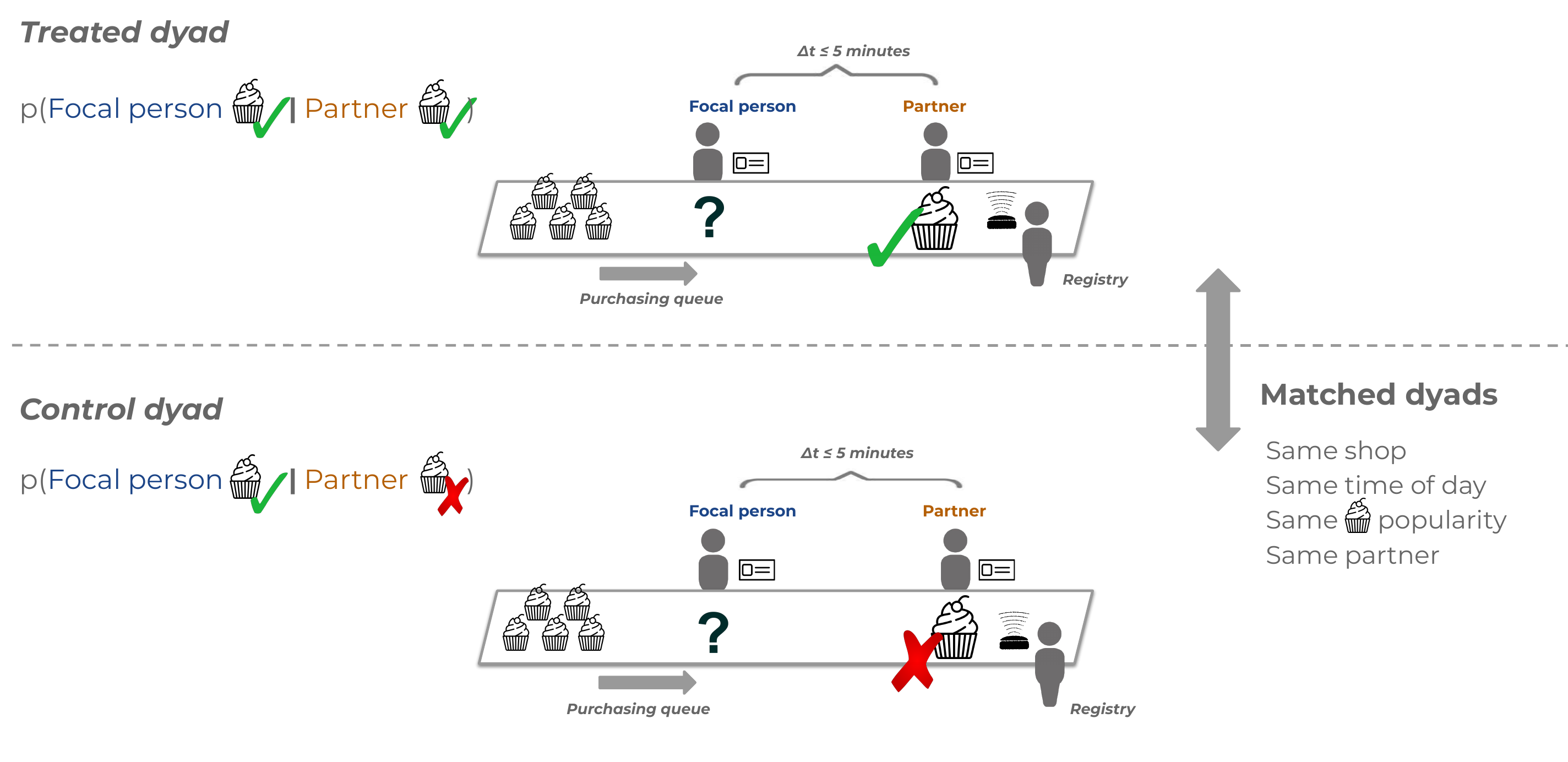}
\caption{\textbf{Study design}. We identify dyads where two individuals make purchases within five minutes of each other, with no one in between, adjacent in the purchasing queue. The first person to make the transaction in the queue is referred to as the partner, and the second person as the focal person. We are interested in identifying the impact that the purchasing behavior of the partner has on the focal person, \eg, purchasing a dessert as in the illustration. To that end, the dyads are matched, such that the dyads are comparable (\ie, they occur in the same shop, time of day, same partner identity, same availability, and popularity of the dessert), but, in treated dyads, the partner purchases a dessert, whereas in the control dyads, the partner does not purchase it. Our study then contrasts the focal person's probability of purchasing the dessert, given that the partner purchased (treated) or not (control).}
\label{fig:study}
\end{figure}

% What we analyze and how, what are the key components of our approach
Based on the transactional data, we design an observational study to identify and measure mimicry in food purchases. We leverage the sequential nature of shop queues and the fact that, with passively sensed data, we can observe many persons in many situations. We consider a large number of adjacent purchases where a focal user immediately follows another user (``partner'') in the checkout queue and both make a purchase. We identify
about 500,000
% 0.5M
such \emph{dyads} (adjacent purchases made by the focal-partner pair) (\cf\ \Figref{fig:study}). The large number of dyads, rich data about the environmental context, and information about the individuals' historical patterns let us make measurements of high granularity and scale. We use a matching\hyp based methodology to identify comparable dyads and identify the effect, while minimizing the impact of biasing factors.

%Benefits of our approach and how it differs from what was done before/improvements, contributions over existing knowledge

\xhdr{Summary of main findings} Analyzing purchasing behaviors, we find significant evidence of mimicry, with partners' purchases affecting all food types. Across food additions purchased during lunchtime together with a meal, we find that the focal user is significantly more likely to purchase the food item if the partner has already bought the item, \vs when the partner has not. The partner's choice to purchase an item increases the focal user's purchasing probability, by {14\% in absolute terms}, or by {83\% in relative terms}. The largest increase in purchasing probability occurs for condiments, while the smallest occurs for soft drinks.

Furthermore, we find that this effect diminishes when we measure the influence of a random (rather than directly preceding) partner on a focal user, demonstrating that the observed effect is not an artifact due to other contextual factors (\cf\ \Secref{subsec1}). The observed effect is robust across subpopulations and affects all genders and statuses, while it is the strongest for students and younger persons (\cf \Secref{subsec2}). 

Our analyses of purchase logs provide novel insights into purchasing mimicry. First, the novel dataset, its scale, and a large number of studied dyads make it possible to study mimicry with greater statistical power compared to previous research.

Second, we perform causal analyses. Given the available information about individuals, who can be consistently observed across many adjacent purchases, and about the environment, such as the popularity and availability of different foods at shops in time, we can minimize the impact of numerous important confounding factors and isolate the mimicry in purchases. We carefully select suitable dyads, aiming to disentangle homophily from influence. Additionally, we quantify the strength of unobservable biases through sensitivity analysis and perform an array of robustness tests.

Third, having access to a multi\hyp year history of all transactions made on a large campus allows us to measure a wide set of purchasing behaviors that occur in the real world, as opposed to the artificial setting of lab-based studies, typically focused on a few selected food items \cite{chartrand2013antecedents,tanner2008chameleons}.

The results of this study elucidate the behavioral mechanism of purchasing mimicry and have further implications for the design of policies and interventions, on university campuses and beyond.
\label{sec:intro}

\section{Results}

\subsection{Study design summary}

%Dataset overall info
We start by briefly outlining the study design. Recall that we leverage a large\hyp data set of shop records made on the EPFL university campus. Each food purchase transaction is attributed with the time it took place, information about the location, the cash register where the transaction took place, and the purchased items. For a subset of users, we additionally leverage demographic information: gender, status at the campus (\ie, whether a person is a student, staff member, or ``other'' status, such as a visitor), and birth year (statistics about the dataset are outlined in Methods, \Secref{data}). We additionally estimate demographic information for the whole population using the paradigm of amplified asking, \ie, by fitting a statistical model to a small subsample with known demographics and applying the model to the remaining population in order to approximately estimate their demographics (Methods, \Secref{aplified_asking}).

%Sequential setup
The study design is illustrated in \Figref{fig:study}. We identify purchases where individuals make purchases within five minutes of each other, adjacent in the queue, with no one in between (referred to as \emph{dyads}). The first person to make the transaction in the queue is referred to as the partner and the second person as the focal person. We are interested in identifying the impact that the purchasing behavior of the partner has on the focal person, \ie, the change in the probability that the focal person will buy a certain food item when the partner buys the same item before the focal person, compared to when the partner does not buy that item. We study dyads where the partner and the focal person are observed together repeatedly (Methods, \Secref{mathods:situations}).

\begin{figure}[t!]
    \begin{minipage}[t]{0.42\textwidth}
    \centering
    \includegraphics[width=\textwidth]{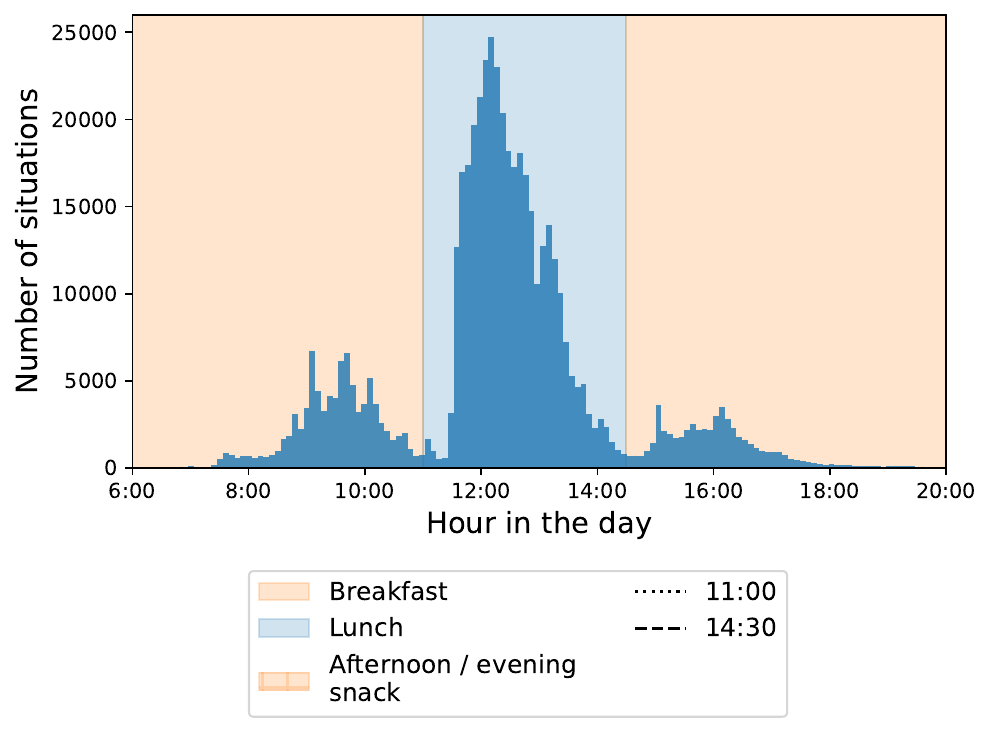}
    \subcaption{}
    \label{fig:1a}
    \end{minipage}
    \hfill
    \begin{minipage}[t]{.52\textwidth}
    \centering
    \includegraphics[width = \textwidth]{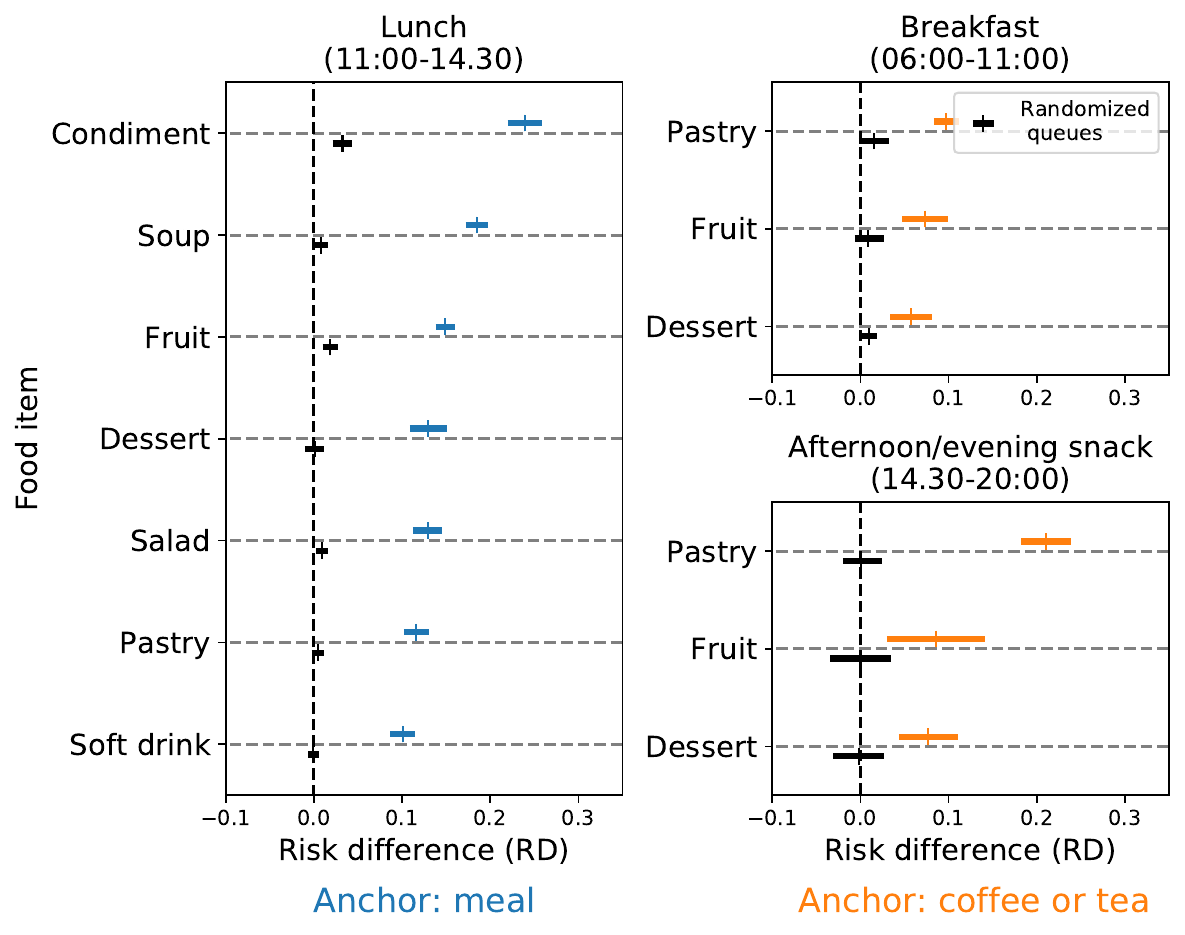}
    \subcaption{}
    \label{fig:1b}
    \end{minipage}
    
    \centering
    \begin{minipage}[t]{0.31\textwidth}
    \centering
    \includegraphics[width=\textwidth]{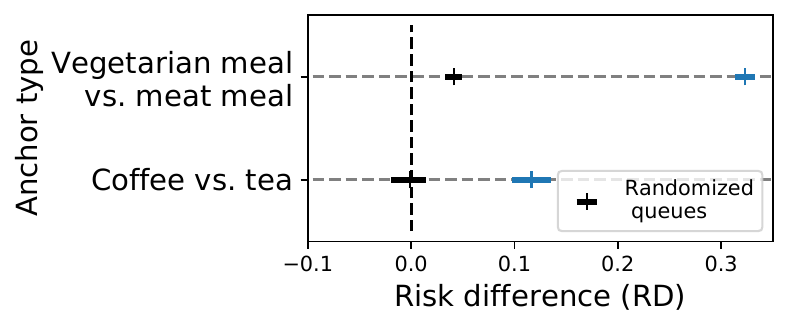}
    \subcaption{}
    \label{fig:anchors}
    \end{minipage}
\caption{\textbf{Purchasing mimicry across times of the day and the food items.} In (a), the histogram of dyads during a day, on the x\hyp axis the hours in the day, and on the y-axis, the number of dyads. The three peaks correspond to breakfast, lunchtime, and afternoon or evening snack time (shaded regions). In (b), separately for lunch, breakfast, and afternoon or evening snack, the estimated risk difference (on the x\hyp axis), for the different food item additions (on the y\hyp axis). In (c), the estimated risk difference (on the x\hyp axis), for the anchor type (on the y\hyp axis), type of meal, vegetarian \vs not, and type of beverage, coffee \vs tea. The error bars mark 95\% bootstrapped CI. Risk difference estimates are colored (blue for lunch where the anchor is the meal, orange for breakfast and afternoon or evening snack where the anchor is a beverage). The randomized baseline is presented in black.
}
\label{fig:1}
\end{figure}

%Anchors setup
The shops typically open at 07:00 and close at 18:00. The studied dyads occur during breakfast (06:00--11:00), during lunch (11:00--14:30), or in the afternoon (14:30--20:00). During the three periods, persons purchase an \textit{anchor}---a meal during lunch or a beverage (coffee or tea) during breakfast or afternoon (\Figref{fig:1a}). In addition to the anchor food item, individuals might purchase an additional item (such as a dessert or a condiment), referred to as an \emph{addition}. In our main analyses, we study the effect of purchasing mimicry of 13 frequent additions (the selection of the food items in focus is outlined in Methods, \Secref{mathods:situations}). We ensure that in the dyads the partner and focal person both purchase an anchor item (a meal during lunchtime or a hot beverage during the morning or afternoon/evening), and observe the purchasing of one of the 13 food additions purchased with the anchor.

%Assumptions and DAG
In this setting, the observed behavior of the focal person is impacted by the partner's traits through their social tie, and by the partner's food choice through the sequential ordering in the queue. Additionally, both the partner's and the focal person's food choice is influenced by common environmental factors. The setting is informed by standard assumptions made to identify the causal effect of social influence under the presence of homophily in a pairwise setup, when examining the causes behind why a person manifested a behavior at a given time \cite{liotsiou2016social,shalizi2011homophily}. The statistical assumptions and the causal graph reflecting them are detailed in Methods (\Secref{sec:dag}).

The minimum sufficient set of variables to control for (according to the backdoor criteria, \cf\ \Figref{fig:dag1}) are the common environmental factors that day (shop, time of day [breakfast time, lunchtime, afternoon], popularity, and availability of the item that day) and the partner's identity, which captures the partner's eating profile.%
\footnote{In Supplementary Material \Secref{sec:alternativedags}, we examine the robustness of our estimates when allowing for violations of these assumptions.}  

%Matching
\xhdr{Estimation} We then perform matching of dyads such that the dyads are comparable (\cf\ Methods,~\Secref{matched_framework}). In the treated dyads in a matched pair, the partner purchases a food item of interest, whereas in the control dyad in the pair, the partner does not purchase the food item of interest. The outcome of the matching is are \emph{matched pairs of dyads}.

After matching, within the matched pairs of comparable dyads where one of the 13 food items is bought or not, we contrast the focal person's probability of purchasing the food item of interest when the partner purchased the item (treated condition) to the probability when the partner did not purchase the item (control condition). The discrepancy between the two probabilities is then expressed in absolute and relative terms using risk difference and risk ratio, respectively (Methods, \Secref{matched_framework}).

%Randomized baseline
\xhdr{Randomized baseline}
Moreover, we consider a randomized baseline. In each dyad, instead of the partner, we choose a random person from the same queue, on the same day, at the same time of day (breakfast, lunch, dinner). The objective of the randomized baseline is to understand similarities stemming from the contextual factors and not directly caused by the actual ordering of the queue and the partner's choice. The estimation, as previously described, is then performed by dyad matching after the queue randomization.

\begin{table}[b!]
\small
\centering
  \caption{\textbf{Contingency table.} The number of matched dyad pairs in each condition (treated and control). In columns, dyads where partner purchased the item, and in rows, matched dyads where partner did not purchase the item.}
  \label{tab:contingency}
\begin{tabular}{c r r r | c}
    \multicolumn{2}{c}{} & \multicolumn{2}{c}{\textit{$\neg$Partner purchased (control dyad)}} & \textbf{Total matched } \\
    %\hline
    \multicolumn{2}{c}{} & \multicolumn{1}{c}{\textbf{$\neg$Focal purchased} } & \textbf{ Focal purchased} &  \textbf{pairs of dyads} \\
    %\hline
    \textit{Partner purchased} &   \textbf{$\neg$Focal purchased} & 28111 (57.97\%) &  5221 (10.77\%) & 33332 (68.74\%)\\
    \textit{(treated dyad)} &   \textbf{Focal purchased} & 12119 (24.99\%) & 3042 (6.27\%) & 15161 (31.26\%) \\
    \hline
   \multicolumn{2}{r}{\textbf{Total matched pairs of dyads}}  &  40230 (82.96\%) & 8263 (17.04\%) & {48493 (100\%)} \\
  \end{tabular}
\end{table}

\subsection{Mimicry of partner's purchases affects all food types}\label{subsec1} 

\xhdr{Paired analyses} As a first look into the matched dyads, we test for evidence of purchasing mimicry and aim to identify the effect pooled across food items. The contingency table (Table \ref{tab:contingency}) counts the frequency of the four possible outcomes, comparing matched pairs of dyads where in one dyad the partner buys, and in the other dyad the partner does not buy, the additional food item (\eg, dessert, condiment, fruit, henceforth referred to as an \textit{addition}). Note that the most frequent outcome is that in both matched pairs of dyads, regardless of the partner, both focal users do not buy the addition. The least likely is that in matched pairs of dyads, regardless of the partner, both focal users buy (since purchasing probabilities are in general low, \cf\ Table \ref{tab:frequency}).

In particular, the discordant instances among the matched pairs of dyads are informative, \ie, the off-diagonal entries in the contingency table, which correspond to matched pairs of dyads where the two focal persons' purchases differ. If there were no partner effects, the two types of discordant entries would be balanced. However, we observe that focal persons mirror their partners more frequently than they do the opposite (2.3 times more likely). In 25\% of matched pairs of dyads, focal persons purchase when partners do and focal persons do not purchase when partners do not. In contrast, the opposite scenario (focal persons doing the opposite of their partners) is rarer, occurring in only 11\% of matched pairs of dyads. The imbalance between the discordant instances serves as first evidence of mimicry. Based on the contingency table, we reject the null hypothesis of no treatment effect ($p<10^{-12}$ according to $\chi^2$-test of no treatment effect).

\xhdr{Risk analyses} Next, pooling the matched pairs of dyads across the different items, we quantify risk difference (RD) and risk ratio (RR) (\cf\ Methods, \Secref{matched_framework}), which serve as the main outcomes in our analyses. Overall, across all matched pairs of dyads (13 food item additions, \eg, condiment, dessert), we find a risk difference of 14.22\% [13.73\%, 14.74\%] and a risk ratio of 1.83 [1.79, 1.88], meaning that the partner's choice to purchase an item increases the focal person's own purchasing probability by 14.22 percentage points in absolute terms, or by 83.48\% in relative terms. In comparison, in the case of the randomized baseline where the purchasing order in the queue is randomized, we find a risk difference of 1.07\% [0.69\%, 1.45\%] and a risk ratio of 1.07 [1.05, 1.1]. In other words, the partner's influence on the focal person nearly entirely disappears once the ordering of the queue is randomized. The gap between true and randomized queues is observed consistently across the nine years spanned by the dataset (Supplementary Material, \Figref{fig:byyear}).

\xhdr{Risk analyses across food items} Since effect modification is expected, for the different times of day and across the 13 additions (seven lunch additions, three breakfast additions, three afternoon/evening snack additions), we quantify the risk differences separately. We find that all the risk differences are significantly different from zero at 95\% confidence level (\Figref{fig:1b}). The random baseline is much smaller for all additions, among different times of day and among additions.
The risk difference for lunch additions varies between 10.06\% [8.65\%, 11.42\%] for soft drinks and 23.94\% [22.11\%, 25.76\%] for condiments. Risk differences for breakfast additions are 5.78\% [3.39\%, 8.37\%] for dessert, 7.34\% for fruit [4.73\%, 9.95\%], and 9.74\% for pastry [8.25\%, 11.18\%]. For afternoon or evening snack, the risk differences are 7.61\% for dessert [4.35\%, 11.02\%], 8.58\% fruit [3.85\%, 14.16\%], and 21.06\% for pastry [18.22\%, 23.89\%]. The relative version of these findings (measured by relative risk) is presented in the Supplementary Materials (\Figref{fig:relative}).

\xhdr{Risk analyses across anchors} Although our main analyses focus on food addition items, we also analyze the mimicry of the anchor itself, within the matched pairs of dyads (\Figref{fig:anchors}). Here the meal anchor comparable dyads can be vegetarian or meat-based, whereas the beverage anchor can be coffee or tea. We observe significant risk differences for meal type (32.28\% [31.39\%, 33.21\%]) and beverage type (11.65\% [9.72\%, 13.52\%]). The randomized baseline is again much lower in both cases (meal type: 4.14\% [3.30\%, 4.95\%]; beverage type: 0.10\% [$-$0.02\%, 0.01\%]). We suspect that mimicry is stronger for the meal-type anchor because purchasing vegetarian food is a behavior related to health and sustainability and, therefore, potentially more likely to be impacted by social norms~\cite{segovia2019health}. We also performed a robustness test requiring that the matched pairs of dyads contain exactly the same anchor (meal \vs\ vegetarian mean; coffee \vs\ tea), described in the Supplementary Material (\Secref{further}) and leading to similar findings as above.

To summarize, among the matched pairs of dyads, we find significant mimicry of partners' purchases affecting all food types. The partner's influence on the focal person diminishes once the ordering of the queue is randomized.

\begin{figure}[t!]
    \centering
     \begin{minipage}[b]{0.38\textwidth}
        \centering
        \includegraphics[width=\textwidth]{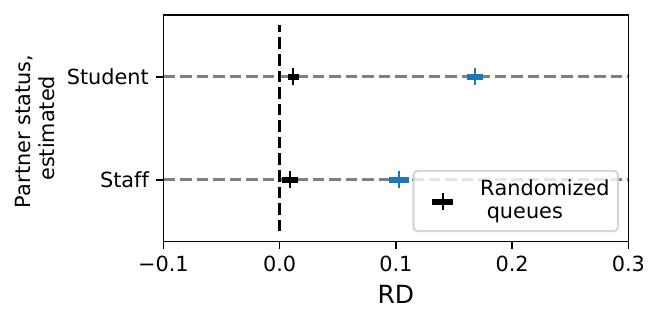}
        \subcaption{}\par
         \label{fig:2a}
        \vspace{\baselineskip}
        \includegraphics[width=\textwidth]{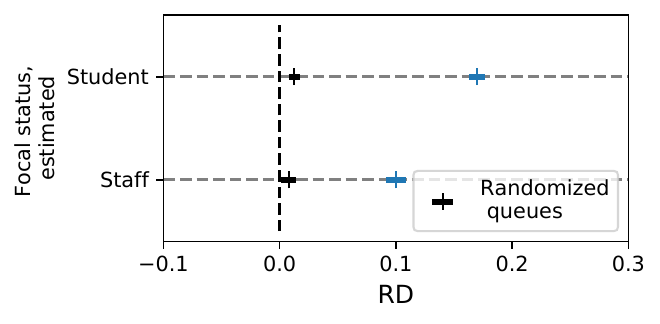}
        \subcaption{}
         \label{fig:2b}
    \end{minipage}
    \begin{minipage}[b]{0.48\textwidth}
        \centering
        \includegraphics[width=\textwidth]{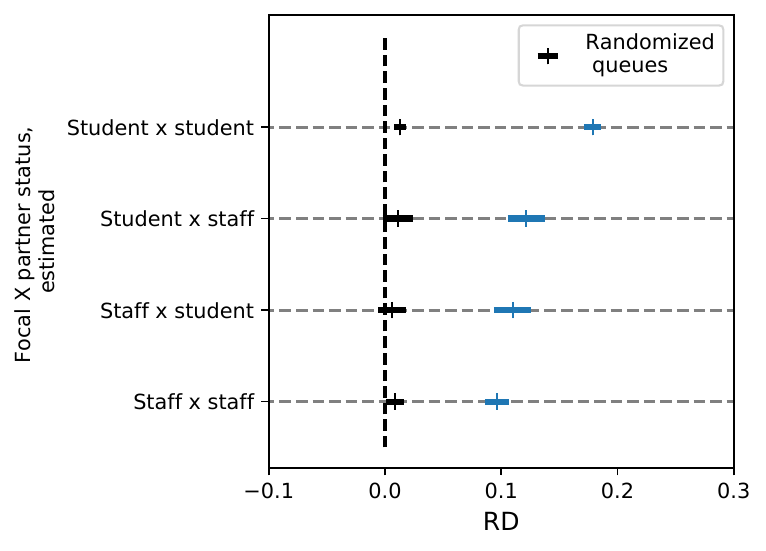}
        \subcaption{}
         \label{fig:2c}
    \end{minipage}

    \caption{\textbf{Effect by the estimated status on campus.} The estimated risk difference across the matched pairs of dyads (on the x\hyp axis), depending on the individuals' estimated status (on the y\hyp axis). The error bars mark 95\% bootstrapped CI. Risk difference estimates are presented in blue, the randomized baseline is presented in black. In (a), for partner's status, in (b), for focal person's status, in (c) for the four combinations of the focal\hyp partner status.
}
\label{fig:2}
\end{figure}

\subsection{Mimicry is strongest among students and the youngest}\label{subsec2} 

\xhdr{Estimated status (students, staff, and other statuses, such as visitors)} We next measure the effect among subsets of matched pairs of dyads based on the estimated status of the partner and the estimated status of the focal person in \Figref{fig:2}. We find that the effect is stronger when the partner is a student (risk difference 16.78\% [16.10\%, 17.46\%]) \vs\ staff member (10.25\% [9.39\%, 11.12\%]; \Figref{fig:2a}). Similarly, the effect is stronger when the focal person is a student (17.0\% [16.30\%, 17.73\%]) \vs\ staff member (10.01\% [9.16\%, 10.91\%]; \Figref{fig:2b}). Examining the four configurations of status within the partner--focal dyad (\Figref{fig:2c}), we find that student--student is the condition with the largest risk difference (17.89\% [17.11\%, 18.60\%]). In contrast, the staff--staff condition has the smallest risk difference (9.66\% [8.60\%, 10.68\%]).

\begin{figure}[t!]
    \centering
    \includegraphics[width=\textwidth]{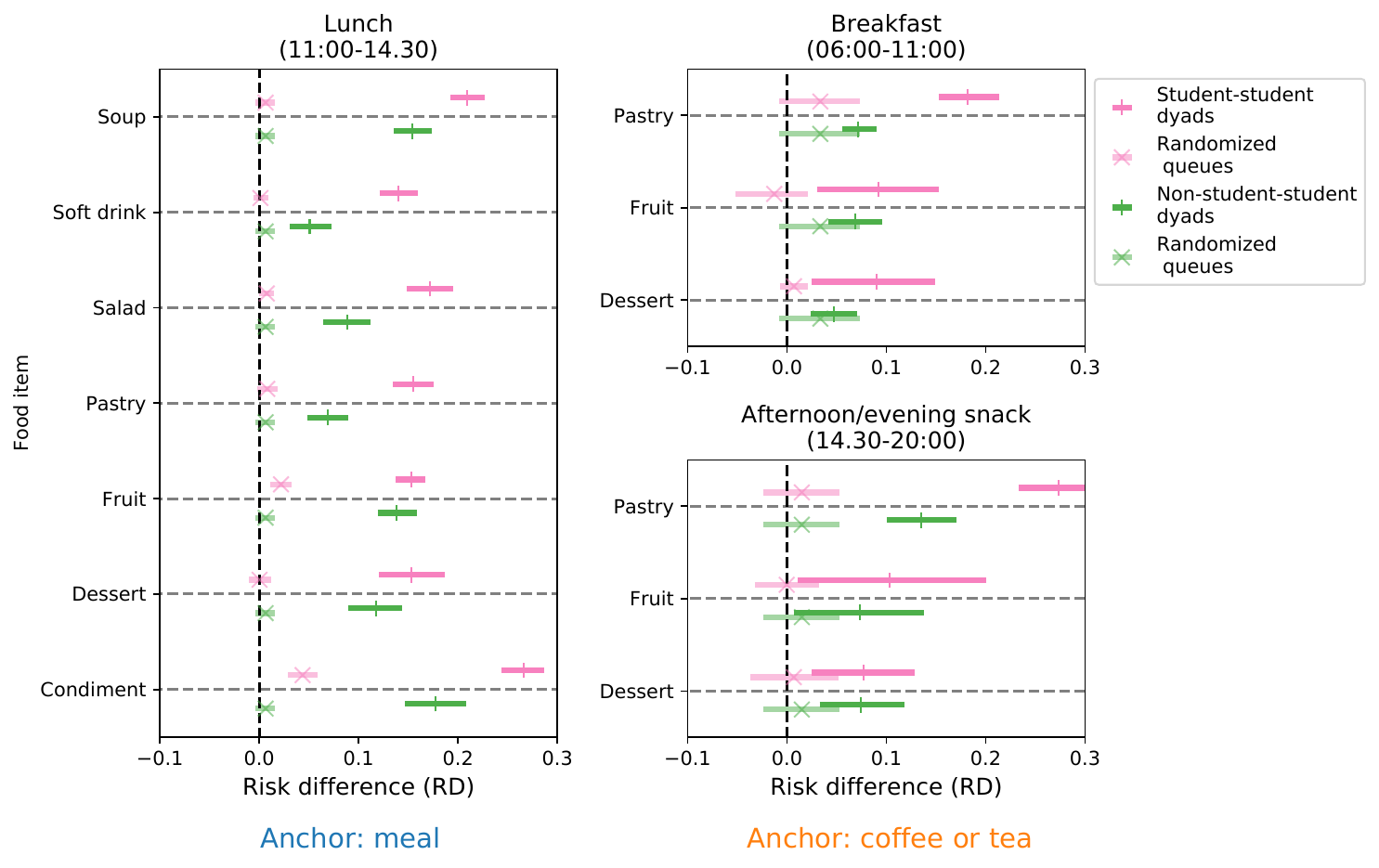}
\caption{\textbf{Effect across dyads, depending on the individuals status on campus.} Separately for lunch, breakfast, and afternoon or evening snack, the estimated risk difference (on the x\hyp axis), for the different food item additions (on the y\hyp axis). The error bars mark 95\% bootstrapped CI. Risk difference estimates are colored in pink for student-student dyads and in green for non-student--student dyads. Randomized baselines are presented in a lighter color.}
\label{fig:statusxfoods}
\end{figure}

The observation regarding students \vs\ staff differences holds across the different foods. In \Figref{fig:statusxfoods}, we measure the risk difference separately among estimated student--student dyads \vs\ all non-student--student dyads where students can be focal or partner, but not both. We find that across the three times of day and the different food items, the effect is consistently greater among the student--student dyads, implying that the difference depending on the status cannot be explained by discrepancies in preferred food items between students and staff. Instead, known moderators of mimicry, including social, emotional, and personality factors, might vary systematically between students and staff and lead to more or less mimicry~\cite{chartrand2013antecedents,tanner2008chameleons}.

\xhdr{Demographics: true status, age, gender} We next investigate the effect across all the matched pairs of dyads within the subpopulation with ground-truth demographic data (Supplementary Material, \Figref{fig:subpop}). First, among the subpopulation with ground-truth status (as opposed to estimated status, as used above; In \Figref{fig:subpopa} and \Figref{fig:subpopb}), we consistently find that the effect is stronger both when the partner is a student (10.73\% [5.67\%, 15.59\%]) \vs\ staff member (5.68\% [1.85\%, 9.38\%]), and when the focal person focal is a student (14.15\% [9.56\%, 19.11\%]) \vs\ staff member (7.22\% [2.44\%, 11.38\%]). Note that the relative ordering is the same as when using estimated status labels. However, the differences are not statistically significant, likely due to the smaller sample size, relative to the above analysis with estimated status labels.

Second, we investigate the role of age (Supplementary Material, \Figref{fig:subpopc} and \Figref{fig:subpopd}). Given the birth date and the time of the transaction, we calculate the age at the time of the transaction, and we bin the age into terciles. We find that the effect is the strongest when both the partner and the focal person are in the youngest group (up to 22 years old at the transaction time). Examining the partner's age, we find that the effect monotonically decreases with age (up to 22 years old: 12.04\% \vs\ 23--32 years old:  8.11\% \vs\ over 32 years old: 4.98\%), and similarly for the focal person's age (up to 22 years old: 17.72\% \vs\ 23--32 years old: 13.18\% \vs\ over 32 years old: 4.04\%).

Third, regarding gender (Supplementary Material, \Figref{fig:subpope} and \Figref{fig:subpopf}), we find significant and similar effects among all subpopulations, with a risk difference greater than zero both when the partner is male and when the partner is female, as well as both when the focal person is male and when the focal person is female. 

To summarize, food choice mimicry is not restricted to particular subpopulations, but observed across all genders, ages, and statuses. The effect is strongest for student-student dyads (\Figref{fig:statusxfoods}) and among younger persons (Supplementary Material, \Figref{fig:subpopc} and \Figref{fig:subpopd}).

\subsection{Mimicry decreases with time lag}\label{subsec3} 

In case of a true causal effect, one would expect a dose--response relationship where the focal person's purchasing probabilities in the matched pairs of dyads diverge more when the two person in the respective dyad are further apart in the queue, as in such cases the focal person is more likely to have seen the choice of the partner.
Hence, we next investigate whether such a dose--response relationship is observed in the data.

\begin{figure}[t!]
    \begin{minipage}[b]{0.32\textwidth}
    \centering
    \includegraphics[width=\textwidth]{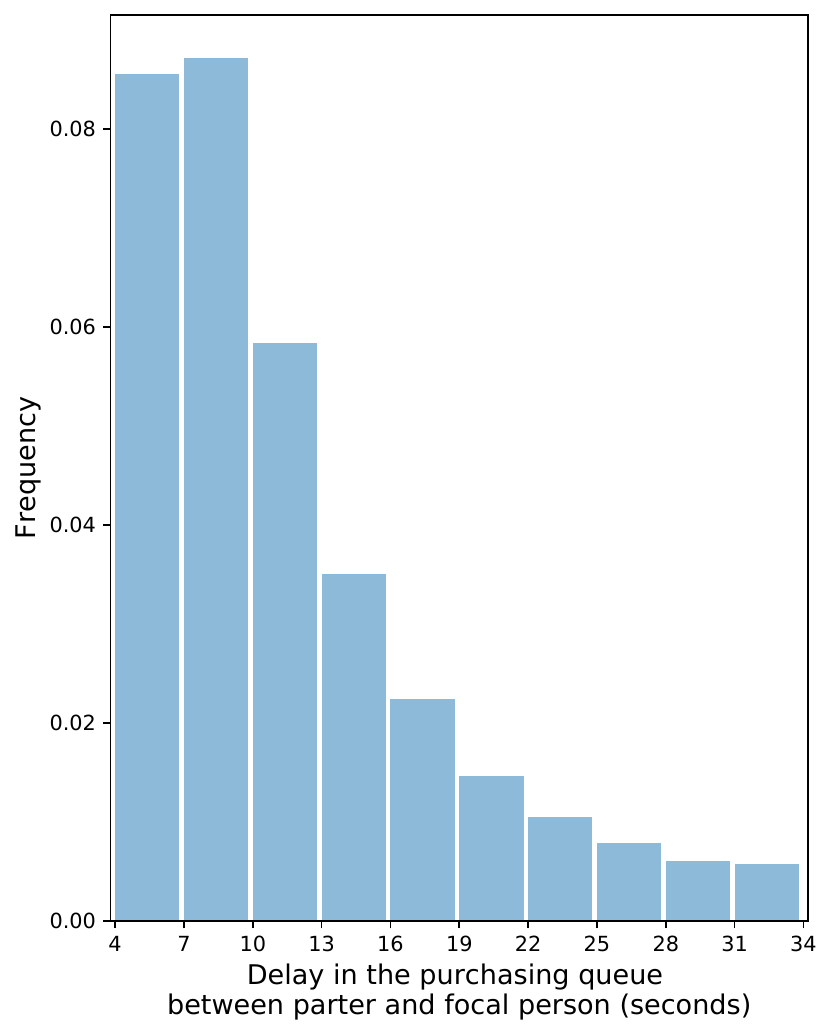}
    \subcaption{Delay histogram}
    \label{fig:dosea}
    \end{minipage}
    \hfill
    \begin{minipage}[b]{.32\textwidth}
    \centering
    \includegraphics[width = \textwidth]{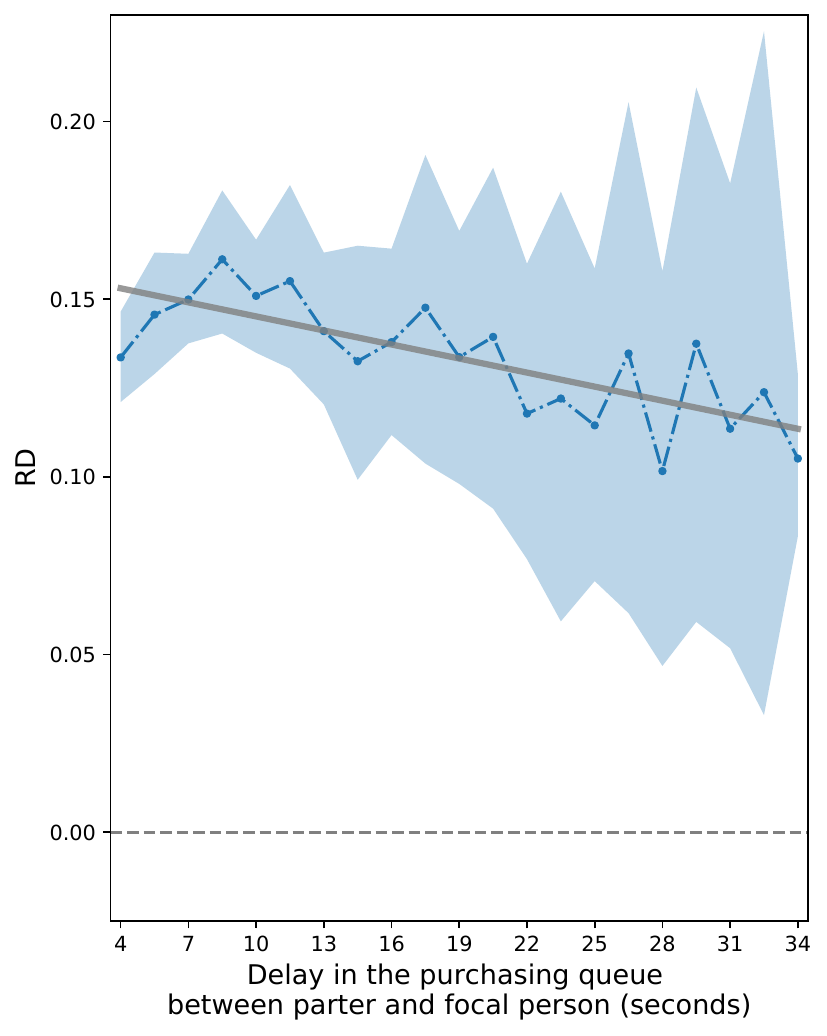}
    \subcaption{Risk difference (RD)}
    \label{fig:doseb}
    \end{minipage}
    \hfill
    \begin{minipage}[b]{.32\textwidth}
    \centering
    \includegraphics[width = \textwidth]{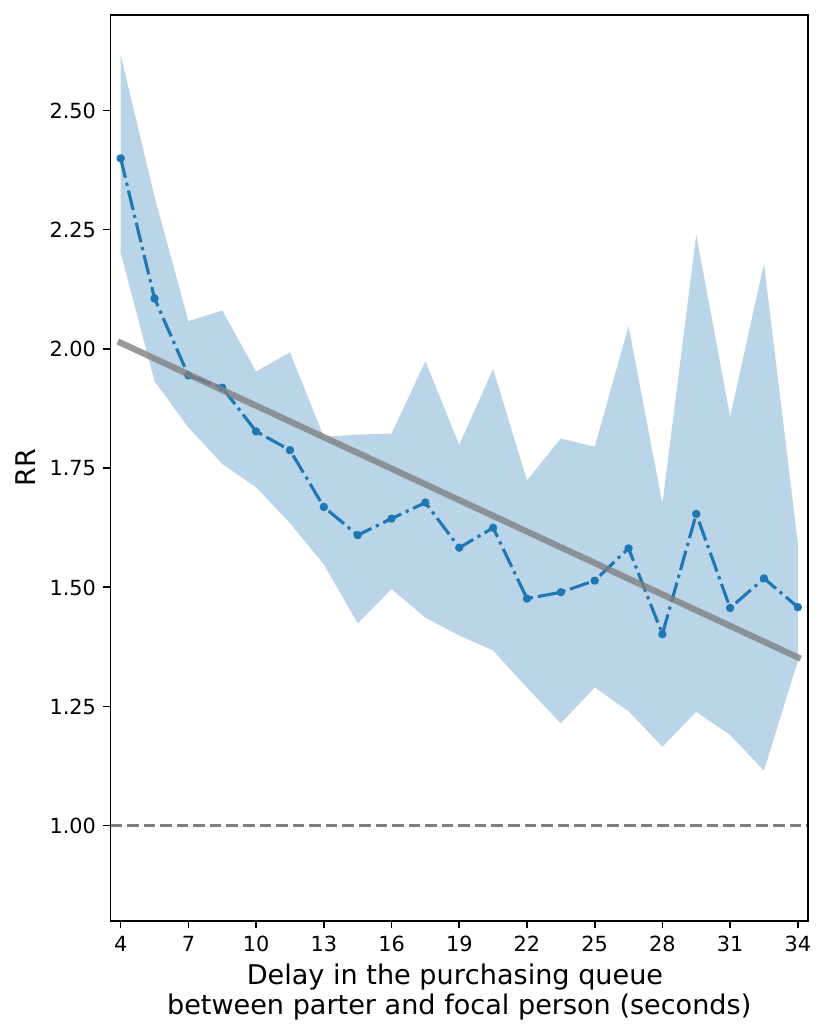}
    \subcaption{Risk ratio (RR)}
    \label{fig:dosec}
    \end{minipage}

    \begin{minipage}[b]{0.18\textwidth}
    \centering
    \includegraphics[width=\textwidth]{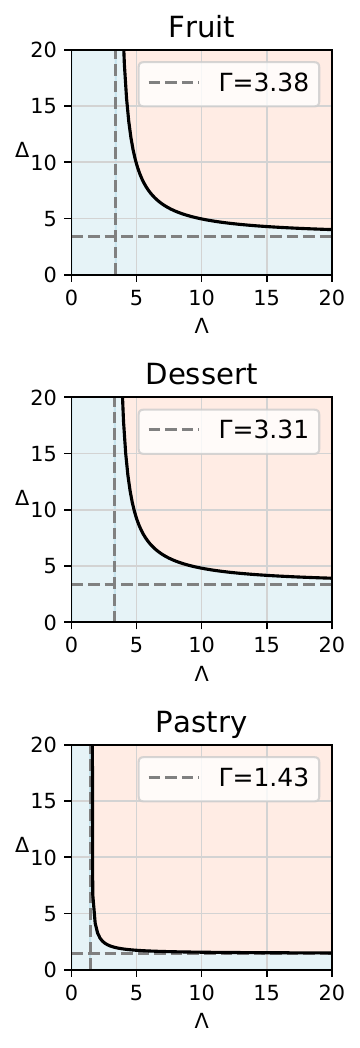}
    \subcaption{Breakfast}
    \label{fig:sensa}
    \end{minipage}
    \hfill
    \begin{minipage}[b]{.18\textwidth}
    \centering
    \includegraphics[width = \textwidth]{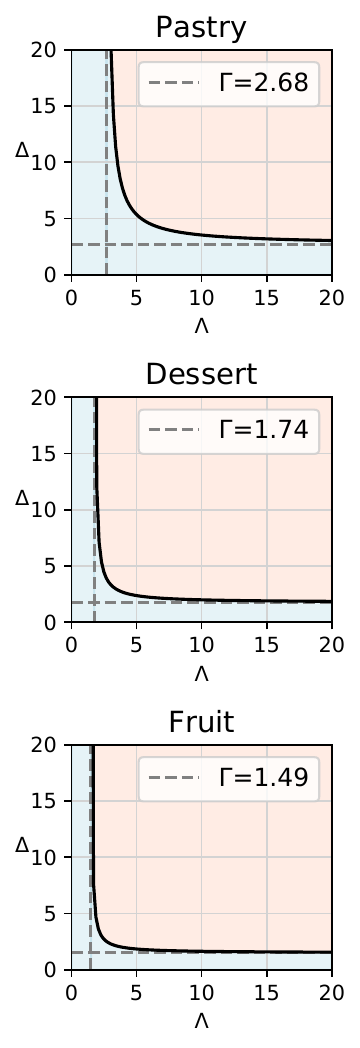}
    \subcaption{Snack}
    \label{fig:sensb}
    \end{minipage}
    \hfill
    \begin{minipage}[b]{.515\textwidth}
    \centering
    \includegraphics[width = \textwidth]{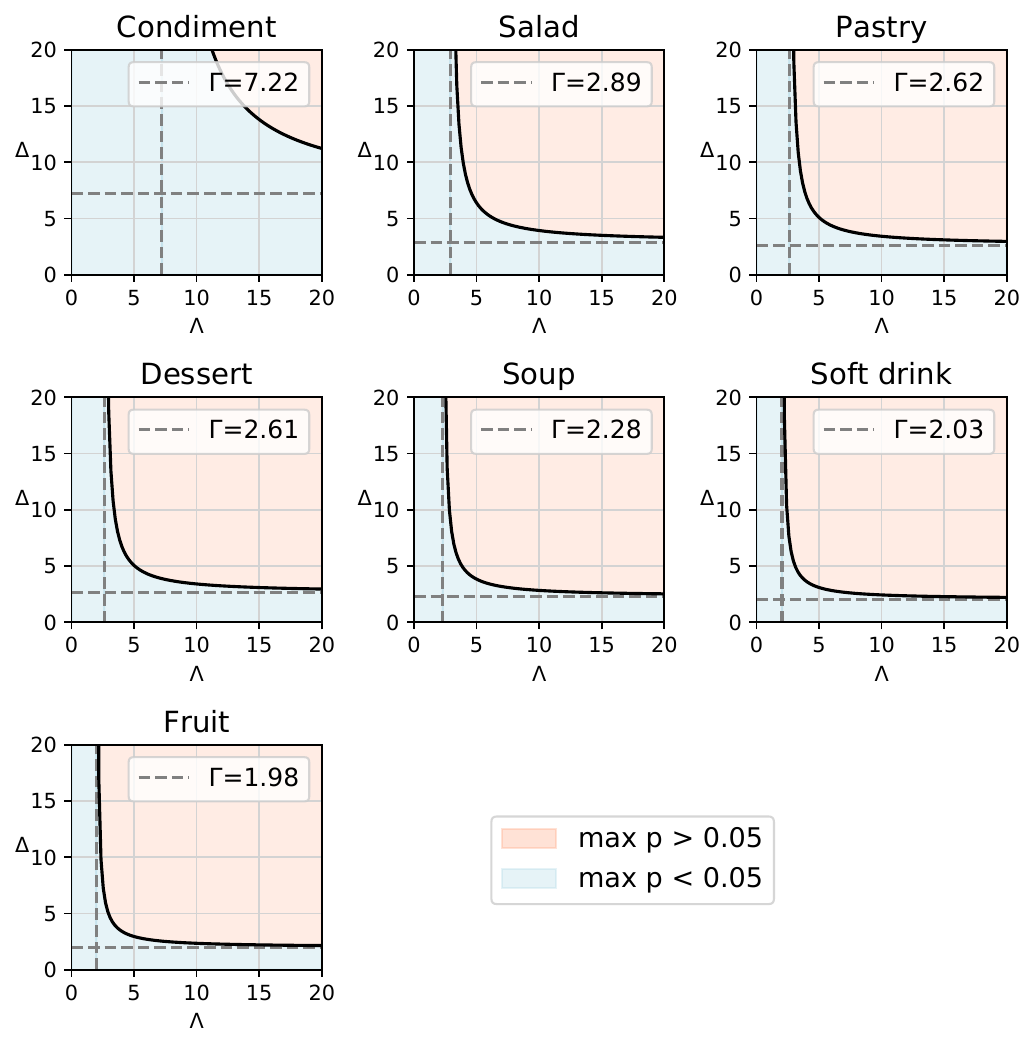}
    \subcaption{Lunch}
    \label{fig:sensc}
    \end{minipage}
\caption{\textbf{Dose\hyp response and sensitivity analysis.} In (a), the histogram of the temporal delay between the partner's and focal person's transactions among the dyads. On the x\hyp axis, the delay, and on the y\hyp axis, the frequency. In (b), the risk difference estimate within the subset of matched pairs of dyads (on the y\hyp axis), with the given delay in the purchasing queue (on the x\hyp axis). In (c), the risk ratio estimate within the subset of matched pairs of dyads (on the y\hyp axis), with the given delay in the purchasing queue (on the x\hyp axis). The shaded areas mark 95\% bootstrapped CI. The gray dashed line represents the least square linear fit. Note the truncated x\hyp axis; dyads with a delay of up to five minutes are considered, however they are rare as visible in (a). In (d), (e), and (f), sensitivity analysis. For the measured sensitivity $\Gamma$, the amplification $(\Lambda,\Delta)$ is plotted. Horizontal and vertical dashed lines indicate $\Gamma$, \ie, the asymptotic value of $\Lambda$ for $\Delta \rightarrow \infty$, and vice versa.
}
\label{fig:dose}
\end{figure}

We find that, as the distance (measured in seconds) between the dyads in the purchasing queue increases (distribution illustrated in \Figref{fig:dosea}), the effect estimate decreases (\Figref{fig:doseb} and \Figref{fig:dosec}). We measure a significant negative association between the delay in the purchasing queue and risk difference (the slope of the linear regression $\beta = -0.002$, two-sided $p = 8.7 \times 10^{-5}$) and between the delay in the purchasing queue and risk ratio ($\beta = -0.03$, two-sided $p = 2.2 \times 10^{-6}$). Overall, a larger effect is observed for smaller distances in the queue.

If other factors were causing the purchasing similarity, such as a third party present in the shop and convincing individuals to purchase a food item or not, and such factors had nothing to do with the ordering and the distance in the purchasing queue, we would not expect to see a dose--response relationship. The latter thus provides further evidence of a causal effect.

\subsection{Robustness tests}\label{sec:sensitivity}

\xhdr{Sensitivity analysis} Our findings rely on the assumption that there are no unobserved variables creating differences between the matched pairs of dyads that could explain the measured purchasing similarity between partners and focal persons (\cf\ Methods, \Secref{sec:dag}). We perform sensitivity analysis to quantify how the estimates made here would change if this assumption were violated to a limited extent. How strong would the unobserved biases need to be to explain the difference in outcomes between the two sets of matched pairs of dyads? Specifically, we measure the following: if there is a violation of the randomized treatment assignment among the matched pairs of dyads (the choice of the partner), how large would it need to be in order to alter the conclusion that the null hypothesis of no differences depending on focal person's choice can be rejected? This quantity is quantified with $\Gamma$, specifying the ratio by which the treatment odds in two matched pairs of dyads would need to differ to result in a $p$-value above the significance threshold (larger values of $\Gamma$ correspond to more robust conclusions).

\Figref{fig:sensa}, \ref{fig:sensb}, and \ref{fig:sensc} summarize the results of the sensitivity analysis. For $p=0.05$, we measure sensitivities $\Gamma$ ranging between 1.43 (purchasing a pastry with a breakfast beverage) and 7.22 (purchasing a condiment with lunch~\footnote{On the studied campus, the condiments are not provided for free and they need to be purchased.}). Additionally, we perform amplification of the sensitivity analysis \cite{rosenbaum2009amplification}, where $\Gamma$ is expressed in terms of two parameters $\Lambda$ and $\Delta$, as $\Gamma = (\Lambda \Delta + 1) / (\Lambda + \Delta)$. Here, $\Delta$ is defined as the strength of the relationship between the unobserved covariate and the difference in outcomes within the matched pair, whereas $\Lambda$ is defined as the strength of the relationship between the unobserved covariate and the treatment assignment.

For combinations of $\Lambda $ and $\Delta$ in the orange area in the figures, significant effects would be detected (leading to $p <0.05$). In contrast, no significant effects would be detected for the combinations in the blue area (leading to $p >0.05$). An infinite number of $(\Lambda,\Delta)$ combinations fall on the border. For instance, in the case of purchasing fruit during breakfast, $(\Lambda, \Delta) = (5.0,9.8)$ corresponds to an unobserved covariate that increases the odds of treatment five-fold and multiplies the odds of a positive pair difference in the outcomes by 9.8. Such amplification is relevant when the concern is not about the violation of randomized treatment assignment but about the presence of specific unobserved covariates with assumed $\Lambda$ or $\Delta$. Overall, we conclude that the study design is insensitive to moderate biases \cite{rosenbaum2018observation}.

\xhdr{Coordination hypothesis} Lastly, we investigated an alternative hypothesis (\cf\ Supplementary Material \Secref{sec:coordination}) where the observed similarities between dyads are driven by the fact that the two persons coordinated to go for a meal together and agreed on the food choice before lining up in the purchasing queue. In such an alternative scenario, people agree in advance, so the order of how they go does not make a difference. However, since we find that the order of how two persons go in the queue does make a difference, we argue that it does not appear plausible that pre-purchase coordination can entirely explain the measured effect.
\label{sec:res}

\section{Discussion}

Since social norms have long been suspected to play a crucial role in the design of dietary behavioral interventions, in this work, our goal is to identify and characterize purchasing mimicry, theorized as the driving mechanism behind the social influences between individuals on campus.

The results presented here document the prominent role of purchasing mimicry and highlight the need for taking it into account when designing dietary interventions and policymaking around how the foods are offered on university campuses and beyond.
First, we find significant mimicry of eating partners' purchases affecting all food types. The partner's influence on the focal person essentially disappears once the ordering of the queue is randomized (\cf\ \Secref{subsec1}). Second, we find that the effect is not restricted to particular subgroups, but is robust across gender, age, and status groups, with the strongest effect sizes for students and younger persons (\cf \Secref{subsec2}). Finally, we find that food choice mimicry decreases with distance in the purchasing queue following a dose--response relationship (\cf\ \Secref{subsec3}).

Overall, we find evidence in favor of a specific behavioral mechanism for how dietary similarities between individuals occur. The evidence is based on a large-scale observational study observing a large number of individuals longitudinally, in a natural setting.

%To promote healthy and sustainable dietary choices in campus environments, stakeholders and commercial food providers need insights regarding food purchasing behaviors on campus. To that end, population\hyp scale transaction data can be leveraged to measure, understand, and enhance purchasing behaviors.

\subsection{Policy implications}

The behavioral mechanism of purchasing mimicry has implications for policies and interventions. Our findings imply that supplementary food items can be leveraged to increase or reduce the intake of specific foods and nutrients via food additions, not only via meals \cite{garnett2019impact}. The fact that we observe discrepancies between subpopulation (\eg, students \vs\ staff) implies that policymakers should take these differences into account when designing food offerings and interventions. Efficiently targeting subpopulations to achieve targeted effects also resonates with previous work, which found that policies aiming to encourage healthy eating need to be optimized for specific subpopulations~\cite{althoff2022large}.

Furthermore, our results indicate that behavioral interventions aiming to change diets should consider leveraging mimicry-based strategies when designing dietary interventions. On the one hand, mimicry can be desirable when the offering is nutritious. On the other hand, mimicry can be undesirable when consuming the affected food items is not recommended. We find that mimicry is strong across the board, for likely healthy items (such as fruit and salad) as well as for potentially less healthy items (such as condiments, desserts, and soft drinks).

While ensuring availability of healthy food and limiting availability of unhealthy food is a general strategy, our study on mimicry suggests interventions to increase the visibility of healthy food choices of others, and decrease the visibility of unhealthy food choices. While previous work has focused on the meals~\cite{garnett2020order}, our findings imply that interventions targeting supplementary food items (such as fruits and desserts, as opposed to meals) are a powerful way to promote the intake of specific foods and nutrients through the mechanisms of social norms.

Concretely, to \emph{promote} purchasing highly nutritious foods in on-campus environments, additions to meals and beverages are a good place to intervene and ensure the availability of fresh, nutritious, and sustainable options and increase the visibility of their purchases. Similarly, to \emph{reduce} purchasing of calorie-dense, low-nutrient foods, additions are a good opportunity for intervening via point-of-purchase intervention strategies \cite{deliens2016dietary}, since they are particularly affected by mimicry (strongest effect is measured for condiments, \cf\ \Figref{fig:1}). Given the goal of reducing individuals influencing each other's purchasing behavior, future work should determine the effectiveness of interventions that aim to reduce detrimental interactions, \eg, by enabling pre-ordering a meal through an application, as opposed to deciding on the spot, since it is known that impulse-buying is mediated by temporal proximity and making decisions in the proximity of others \cite{sharma2010impulse}. Given that mimicry can be good or bad for meal healthiness depending on the specific food items, future work should design interventions that harvest mimicking healthy food purchases while avoiding mimicking unhealthy food purchases.

Similarly, dietary interventions can involve rethinking the design of queuing systems to increase the likelihood that dyads with specific characteristics appear. For instance, a ``bring a student to lunch'' day, where a faculty member takes a student for lunch and is reimbursed if they order a healthy meal, might incentivize specific pairings and corresponding queuing sequences to promote purchases of healthy foods. Since the strongest effects are observed for student-student dyads (\Figref{fig:statusxfoods}), interventions can incentivize social eating with students who purchase nutritious items, by providing them with vouchers to bring a friend to lunch. The question remains whether the impact of naturally occurring social interactions is expected to mirror the impact of externally induced interventions.

Such externally induced interventions should be explored in conjunction with designing dedicated queuing lines to control mixing of people at the check-out registers (\eg, via separate lines for students and staff), manipulating food position (proximity or order) \cite{bucher2016nudging}, since changing the order in which food items are presented at shop counters has been proposed as a potentially effective way of altering food consumption \cite{garnett2020order}. For instance, condiments might be moved away further from checkout areas. We note, however, that the general learnings of this study can equally apply to shops that do not have the the same sequential queueing setup.

Finally, our findings demonstrate that digital traces can provide valuable insights into the determinants of dietary choices. Digital traces can complement small-scale field experiments, making it possible to observe large populations over long periods. By studying behaviors as they occur naturally in a large population, our findings confirm and refine knowledge mostly derived from small-scale experimental studies. Leveraging passively sensed purchase logs makes it possible to anticipate the impact of interventions before implementing them and identify the right subpopulations to target.
For instance, social influence in dietary habits has previously been examined in the context of school children \cite{finnerty2010effects,patrick2005review,salvy2008effects,birch1980effects} and adolescents \cite{stevenson2007adolescents,DELAHAYE2010161,DELAHAYE2011719}, who are theorized to be most susceptible to social pressures to diets and activity patterns \cite{ball2010healthy,salvy2012influence}. Although previous experimental studies found relationship type, gender, and age group not to be significant predictors of eating mimicry \cite{bell2019sensing}, a recurrent issue faced by previous studies is the small sample size. Our findings, by relying on observations with a greater statistical power, confirm and refine the existing knowledge. For instance, we discover the role of age, since we find the effect to be the strongest in the youngest subpopulations and students. 

This latter finding also serves to further emphasize the responsibility that universities have towards their students, extending beyond the key missions of education and research: since universities often provide food on premises, they are also responsible for creating healthy nutritional environments supporting their primary educational goals. In turn, healthy nutritional environments can be achieved by implementing policies and interventions that leverage social determinants of food choice, as previously illustrated in the example interventions to increase the visibility of healthy food choices of others, and decrease the visibility of unhealthy food choices.

\subsection{Limitations}

In terms of limitations, our study examines the behavior of a population situated in Switzerland, a large fraction of which is young and not representative of the global population. Also, the individuals in our population do not exclusively consume foods bought on campus. They may bring food to campus from elsewhere, and they also consume food off campus, implying that food purchase behaviors of our population are only partially observed. Moreover, those who tend to eat food from elsewhere might fundamentally differ from the population that consumes food at campus locations. We note that we study food purchasing, not food consumption \cite{sen2021total,olteanu2019social}, and that it is not guaranteed that the purchased items were consumed by the person who made the transaction. It is also unknown when the purchasing decision is made, since the purchasing decision is only measured through the logged purchasing act. The robustness rest accessing possible pre-purchase coordination and decision-making (\cf\ \Secref{sec:sensitivity}) aims to address this limitation. A further source of measurement error is the fact that the estimation of status is imperfect and that individuals with demographic information might not be representative subpopulation of the complete campus population.

Starting from a set of 16.6M transactions executed in a shop and assigned a person ID, we identified 1M transactions paired into 500,000 dyads executed in close temporal proximity, by people who often eat together. This design choice was made with the goal of studying choices made nearby, by frequent partners, in order to be able to repeatedly observe the same individuals and control for their identity, as necessary to isolate the mimicry effect (\cf\ Methods, \Secref{sec:dag}). However, the identified subset of purchases and the individuals that execute them might not be representative of the complete set of transactions and all the individuals on the campus. Those who eat in close temporal proximity to others might be different from those who only visit shops on less busy occasions and might not exhibit the mimicry patterns described here. For instance, they might be more social, younger, and therefore more susceptible to the choices of others~\cite{krosnick1989aging,ahmed2020susceptibility}. Thus, we can only make claims regarding the studied purchase instances and the observed individuals.

Lastly, we do not have access to fine-grained inventory information used for keeping track of items available at shops at a given time. Therefore, we approximate item availability at the purchase point by identifying what items were bought at least once, rather than via explicit availability information. Therefore, the availability at the purchase point can conceivably vary between dyads in ways that cannot be measured from sales logs alone. Further biases stem from the fact that purchasing behavior and choice mimicry might be driven by other unobserved factors, \eg, purchasing power, personal relationships, overall health and wellbeing, or calorie need. The threat to validity from such unobserved confounds is mitigated by our sensitivity analysis (\cf\ \Secref{sec:sensitivity}), which led us to conclude that the study design is insensitive to small and moderate biases \cite{rosenbaum2018observation}.

\subsection{Future work}

% immediate next directions (1)
This study opens the door for future research directions and potential follow-up studies of the social determinants of food choice. Future work should focus on further understanding what drives the differences between age and status.

% immediate next directions (2)
Moreover, our analyses observe dyads only. Future work should study more complex group dynamics beyond dyads that might take place in purchasing queues. For instance, future work could investigate the presence of a cascading effect in the queue with the partner influencing the focal person, the focal person influencing the next person, and so on. Similarly, future work should determine if there are people who influence often but are rarely influenced themselves, and vice versa.

% bigger picture kind of things (interventions and beyond campus)

Going further, future work should design and deploy on-site interventions to test the potential of behavioral nudges exploiting mimicry to promote healthy and sustainable eating on campuses. Finally, future work should determine the extent to which these results generalize beyond university campus environments, to the general population and further settings where people make food choices while exposed to the choices of others, both acquaintances and strangers, in locations such as supermarkets, coffee shops, bakeries, food courts, food trucks, and so on.

\subsection{Conclusion} The results of this study elucidate the behavioral mechanism of purchasing mimicry and have implications for understanding dietary behaviors on campus. Furthermore, we demonstrated how purchase logs can be leveraged to derive insights into social determinants of dietary behaviors. We hope that this study will inspire other institutions to consider analyzing purchase logs collected as part of regular operations in order to derive insights and design interventions with tangible benefits across communities.

\label{sec:disscussion}

\section{Online Methods}

\subsection{Data}\label{data}

We leverage an anonymized dataset of food purchases made on the EPFL university campus. The data spans from 2010 to 2018, and contains about 18 million transactions made with a badge that allows linking to an anonymized person's ID. The data includes 38.7k users who, on median, are observed for 578 days and make 188 transactions. Each transaction is additionally attributed with the time it took place, information about the location, the cash register where the transaction took place, and the purchased items. The data covers all the food outlets permanently located on campus, including restaurants, cafes, and vending machines. We analyze adjacent purchases (referred to as \emph{dyads}) made in one of the twelve major catered shops (as opposed to self-service vending machines, \ie 16.6M transactions in total). The shops are illustrated in the Supplementary material, \Secref{layouts}. Furthermore, food items are associated with unstructured textual descriptions. The unstructured textual descriptions were additionally mapped to categorical labels such as ``meal'' or ``dessert'' by a research assistant \cite{gligoric2021ties}.

We also rely on a smaller-size enriched transactional dataset gathered during a three-week campus-wide sustainability challenge in November 2018, during which 1,031 consenting participants formed teams to compete in taking sustainable actions. For this subset of users, we leverage demographic information: gender (584 female, 447 male), status at the campus (724 students, 280 staff, 27 other), and birth year (average 1991, median 1994, Q1 1988, Q3 1998).

\subsection{Sequential choices: studied dyads}\label{mathods:situations}

\begin{table}[b!]
  \caption{\textbf{Food addition item statistics.} For the three meals, the studied food addition items, the frequency with which the addition is purchased by the partner within the studied dyads (\ie, treatment frequency), and the number of matched pairs of dyads where the addition is purchased \vs\ not. }
  \label{tab:frequency}
 \centering 
\begin{tabular}{llrr}
\toprule
Time of day & Food addition item & Treatment freq. & \# matched pairs of dyads  \\
\midrule

Breakfast/morning snack time & Dessert &  8.77\% &  1004 \\
      & Fruit &   3.62\% & 1226 \\
      & Pastry &  7.85\% &16898 \\
      
Lunch time & Condiment &  1.49\% &  5590 \\
      & Dessert &  1.21\% & 3954 \\
      & Fruit &   8.49\% & 22424 \\
      & Pastry &   1.93\% & 7400  \\
      & Salad &   1.51\% & 5286 \\
      & Soft drink &  2.8\% & 8970 \\
      & Soup &  7.79\% &  18956 \\
      
Afternoon/evening snack time & Dessert &  6.39\% &   1288 \\
      & Fruit &   2.7\% & 466 \\
      & Pastry &  16.54\% &  3524 \\
\bottomrule
\end{tabular}
\end{table}

In order to identify purchasing mimicry, we observe a sequence of transactions made using staff or student badges in the queue of a cash registry, in a given shop. We identify instances when two individuals are adjacent in the queue and make a transaction within five minutes of each other, with no one between them. We observe two individuals making a purchase sequentially in a purchasing queue with the badge, as illustrated in \Figref{fig:study}.

Co-purchasing matrices (Supplementary Material, Tables \ref{tab:3}, \ref{tab:2}, and \ref{tab:1}) outline the dyad frequency among the subset of the studied dyads with demographic data available. The tables illustrate a preference for eating with others of the same gender, age, and status. We also note that the order female-male is more common than the order male-female. Similarly, the order staff-student is more common than the order student-staff, likely reflecting social norms of politeness and giving way to others depending on their gender and seniority.

A unit of analysis is an instance of two persons having a meal together (a dyad), operationalized as two individuals executing transactions consecutively in the same shop, at the same purchasing line, on the same day, within a 5-minute window, with no one else executing a transaction in between. 

There are three daily three peaks of transactions. The studied dyads occur during the time of breakfast (06:00--11:00), lunch (11:00--14.30), or afternoon (14:30--20:00). During the three periods, persons purchase an anchor---a meal during lunch or a beverage (coffee or tea) during breakfast or afternoon (\Figref{fig:1a}). Coffee is a more frequent anchor compared to tea. During breakfast, tea is an anchor in 12.99\% of dyads, \vs coffee in 87.01\%. During afternoon, tea an is anchor in 14.05\% of dyads, \vs coffee in 85.95\%.

In addition to the anchor food item, individuals might purchase an additional item (such as a dessert or a condiment), referred to as \emph{an addition}. In our main analyses, we study the effect of purchasing mimicry of the frequent additions. The additions were selected to include all food items where among the dyads with the anchor, in at least 1\% of dyads, the partner buys the addition (Table \ref{tab:frequency}) (\ie, at least 1\% of the dyads is treated). In total, there are three types of additions frequently purchased together with a beverage during breakfast and afternoon hours (fruit, dessert, and pastry), and seven types of additions frequently purchased together with a meal during lunch hours (condiment, salad, pastry, dessert, soup, soft drink, and fruit). Note that pastry is a separate category from dessert since it can be savory. 

Overall, we analyze $509,220$ identified dyads that took place in one of the twelve major shops. The $509,220$ dyads are executed by $18,494$ unique individuals. The instances are selected such that the two individuals make at least ten transactions together adjacent in the purchasing queues in order to be able to observe the same pairs repeatedly, at least ten times. The threshold was selected to be able to repeatedly observe the same individuals and control for their identity, as necessary to isolate the mimicry effect (\cf\ \Secref{sec:dag}). In a supplementary analysis, we examine the impact that the number of adjacent transactions has on the estimated effect size (Supplementary materials, \Secref{sec:alternativedags}).

\subsection{Causal assumptions and Directed Acyclic Graph (DAG)}\label{sec:dag}

\begin{figure}[t!]
\centering
\begin{subfigure}[b]{0.4\linewidth}
\centering

\begin{tikzpicture}
% nodes %
\node[text centered] (a) {$X_a$};
\node[below=of a, xshift=1cm, text centered] (b) {$S_{a,b}$};
\node[right=of a, xshift =1cm, text centered] (c) {$X_b$};
\node[below=of a, yshift=-1cm, xshift=0cm, text centered] (d) {$Y_a(t)$};
\node[below=of a, yshift=-1cm, xshift=2.2cm, text centered] (e) {$Y_b(t)$};
\node[below=of b, yshift=-1cm, text centered] (f) {$P(t)$};
% edges %
\draw[-{Stealth[slant=0]}, line width= 1,] (a) -- node[above,font=\footnotesize]{}  (b);
\draw[-{Stealth[slant=0]}, line width= 1,] (c) -- node[above,font=\footnotesize]{}  (b);
\draw[-{Stealth[slant=0]}, line width= 1,] (b) -- node[above,font=\footnotesize]{}  (e);
\draw[purple, -{Stealth[slant=0]}, line width= 1,] (d) -- node[text width=2cm, above,font=\tiny]{}  (e);
%\draw[-{Stealth[slant=0]}, line width= 1,, gray] (b) -- node[above,font=\footnotesize]{}  (d);
\draw[-{Stealth[slant=0]}, line width= 1,] (f) -- node[above,font=\footnotesize]{}  (d);
\draw[-{Stealth[slant=0]}, line width= 1,] (f) -- node[above,font=\footnotesize]{}  (e);
\draw[-{Stealth[slant=0]}, line width= 1,] (a) -- node[above,font=\footnotesize]{}  (d);
\draw[-{Stealth[slant=0]}, line width= 1,] (c) -- node[above,font=\footnotesize]{}  (e);
%\draw[-{Stealth[slant=0]}, line width= 1, gray] (a) to  [out=45,in=45, looseness=2] node[below, font=\footnotesize]{} (e);
%\draw[-{Stealth[slant=0]}, line width= 1, gray] (c) to  [out=135,in=135, looseness=2] node[below, font=\footnotesize]{} (d);

%\draw[->, line width=1] (t) to  [out=270,in=270, looseness=0.3] node[below, font=\footnotesize]{$\delta$} (y);
\end{tikzpicture}
\subcaption{}
\label{fig:dag1}
\end{subfigure}
~
\begin{subfigure}[b]{0.4\linewidth}
\centering

\begin{tikzpicture}
% nodes %
\node[text centered] (a) {$X_a$};
\node[below=of a, xshift=1cm, text centered] (b) {$S_{a,b}$};
\node[right=of a, xshift =1cm, text centered] (c) {$X_b$};
\node[below=of a, yshift=-1cm, xshift=0cm, text centered] (d) {$Y_a(t)$};
\node[below=of a, yshift=-1cm, xshift=2.2cm, text centered] (e) {$Y_b(t)$};
\node[below=of b, yshift=-1cm, text centered] (f) {$P(t)$};

% edges %
\draw[-{Stealth[slant=0]}, line width= 1, gray, dashed] (a) to  [out=45,in=45, looseness=1.7] node[below, font=\footnotesize]{} (e);
\draw[-{Stealth[slant=0]}, line width= 1,] (a) -- node[above,font=\footnotesize]{}  (b);
\draw[-{Stealth[slant=0]}, line width= 1,] (c) -- node[above,font=\footnotesize]{}  (b);
\draw[-{Stealth[slant=0]}, line width= 1,] (b) -- node[above,font=\footnotesize]{}  (e);
\draw[purple, -{Stealth[slant=0]}, line width= 1,] (d) -- node[above,font=\footnotesize]{}  (e);
%\draw[-{Stealth[slant=0]}, line width= 1,, gray] (b) -- node[above,font=\footnotesize]{}  (d);
\draw[-{Stealth[slant=0]}, line width= 1,] (f) -- node[above,font=\footnotesize]{}  (d);
\draw[-{Stealth[slant=0]}, line width= 1,] (f) -- node[above,font=\footnotesize]{}  (e);
\draw[-{Stealth[slant=0]}, line width= 1,] (a) -- node[above,font=\footnotesize]{}  (d);
\draw[-{Stealth[slant=0]}, line width= 1,] (c) -- node[above,font=\footnotesize]{}  (e);
%\draw[-{Stealth[slant=0]}, line width= 1, gray] (a) to  [out=45,in=45, looseness=2] node[below, font=\footnotesize]{} (e);
%\draw[-{Stealth[slant=0]}, line width= 1, gray] (c) to  [out=135,in=135, looseness=2] node[below, font=\footnotesize]{} (d);

%\draw[->, line width=1] (t) to  [out=270,in=270, looseness=0.3] node[below, font=\footnotesize]{$\delta$} (y);
\end{tikzpicture} 
\subcaption{}
\label{fig:dag2}
\end{subfigure}
~

\begin{subfigure}[b]{0.4\textwidth}
\centering
\hspace{-1.5cm}
\begin{tikzpicture}
% nodes %
\node[text centered] (a) {$X_a$};
\node[below=of a, xshift=1cm, text centered] (b) {$S_{a,b}$};
\node[right=of a, xshift =1cm, text centered] (c) {$X_b$};
\node[below=of a, yshift=-1cm, xshift=0cm, text centered] (d) {$Y_a(t)$};
\node[below=of a, yshift=-1cm, xshift=2.2cm, text centered] (e) {$Y_b(t)$};
\node[below=of b, yshift=-1cm, text centered] (f) {$P(t)$};
% edges %
\draw[-{Stealth[slant=0]}, line width= 1,] (a) -- node[above,font=\footnotesize]{}  (b);
\draw[-{Stealth[slant=0]}, line width= 1,] (c) -- node[above,font=\footnotesize]{}  (b);
\draw[-{Stealth[slant=0]}, line width= 1,] (b) -- node[above,font=\footnotesize]{}  (e);
\draw[purple, -{Stealth[slant=0]}, line width= 1,] (d) -- node[above,font=\footnotesize]{}  (e);
%\draw[-{Stealth[slant=0]}, line width= 1,, gray] (b) -- node[above,font=\footnotesize]{}  (d);
\draw[-{Stealth[slant=0]}, line width= 1,] (f) -- node[above,font=\footnotesize]{}  (d);
\draw[-{Stealth[slant=0]}, line width= 1,] (f) -- node[above,font=\footnotesize]{}  (e);
\draw[-{Stealth[slant=0]}, line width= 1,] (a) -- node[above,font=\footnotesize]{}  (d);
\draw[-{Stealth[slant=0]}, line width= 1,] (c) -- node[above,font=\footnotesize]{}  (e);
%\draw[-{Stealth[slant=0]}, line width= 1, gray, dashed] (a) to  [out=45,in=45, looseness=1.7] node[below, font=\footnotesize]{} (e);
\draw[-{Stealth[slant=0]}, line width= 1, gray, dashed] (c) to  [out=135,in=135, looseness=1.7] node[below, font=\footnotesize]{} (d);

%\draw[->, line width=1] (t) to  [out=270,in=270, looseness=0.3] node[below, font=\footnotesize]{$\delta$} (y);
\end{tikzpicture}
\subcaption{}
\label{fig:dag3}
\end{subfigure}
~
\begin{subfigure}[b]{0.4\linewidth}
\centering

\hspace{-1.5cm}
\begin{tikzpicture}
% nodes %
\node[text centered] (a) {$X_a$};
\node[below=of a, xshift=1cm, text centered] (b) {$S_{a,b}$};
\node[right=of a, xshift =1cm, text centered] (c) {$X_b$};
\node[below=of a, yshift=-1cm, xshift=0cm, text centered] (d) {$Y_a(t)$};
\node[below=of a, yshift=-1cm, xshift=2.2cm, text centered] (e) {$Y_b(t)$};
\node[below=of b, yshift=-1cm, text centered] (f) {$P(t)$};
% edges %
\draw[-{Stealth[slant=0]}, line width= 1,] (a) -- node[above,font=\footnotesize]{}  (b);
\draw[-{Stealth[slant=0]}, line width= 1,] (c) -- node[above,font=\footnotesize]{}  (b);
\draw[-{Stealth[slant=0]}, line width= 1,] (b) -- node[above,font=\footnotesize]{}  (e);
\draw[purple, -{Stealth[slant=0]}, line width= 1,] (d) -- node[above,font=\footnotesize]{}  (e);
%\draw[-{Stealth[slant=0]}, line width= 1,, gray] (b) -- node[above,font=\footnotesize]{}  (d);
\draw[-{Stealth[slant=0]}, line width= 1,] (f) -- node[above,font=\footnotesize]{}  (d);
\draw[-{Stealth[slant=0]}, line width= 1,] (f) -- node[above,font=\footnotesize]{}  (e);
\draw[-{Stealth[slant=0]}, line width= 1,] (a) -- node[above,font=\footnotesize]{}  (d);
\draw[-{Stealth[slant=0]}, line width= 1,] (c) -- node[above,font=\footnotesize]{}  (e);
\draw[-{Stealth[slant=0]}, line width= 1, gray, dashed] (a) to  [out=45,in=45, looseness=1.7] node[below, font=\footnotesize]{} (e);
\draw[-{Stealth[slant=0]}, line width= 1, gray, dashed] (c) to  [out=135,in=135, looseness=1.7] node[below, font=\footnotesize]{} (d);

%\draw[->, line width=1] (t) to  [out=270,in=270, looseness=0.3] node[below, font=\footnotesize]{$\delta$} (y);
\end{tikzpicture} 
\subcaption{}
\label{fig:dag4}
\end{subfigure}
\caption{\textbf{Directed acyclic graphs (DAGs)}. DAGs encode the assumptions about the causal relationship between variables. $X_a$ and $X_b$ are partner's and focal person's eating profile respectively; $S_{a,b}$ is the social tie strength; $Y_a(t)$ and $Y_b(t)$  are partner's and focal person's sets of purchased items at time $t$ respectively, and $P(t)$ are common environmental factors at time $t$. Purple arrow marks the causal path of purchasing mimicry. In (a), the assumed DAG. In (b), (c), and (d), the variations of the assumed causal relationships where the Assumption \ref{as:1} is violated such that the traits of the individuals can influence the observed purchasing behavior through factors not related to friendship strength $S_{a,b}$.}
\label{fig:my_label}
\end{figure}
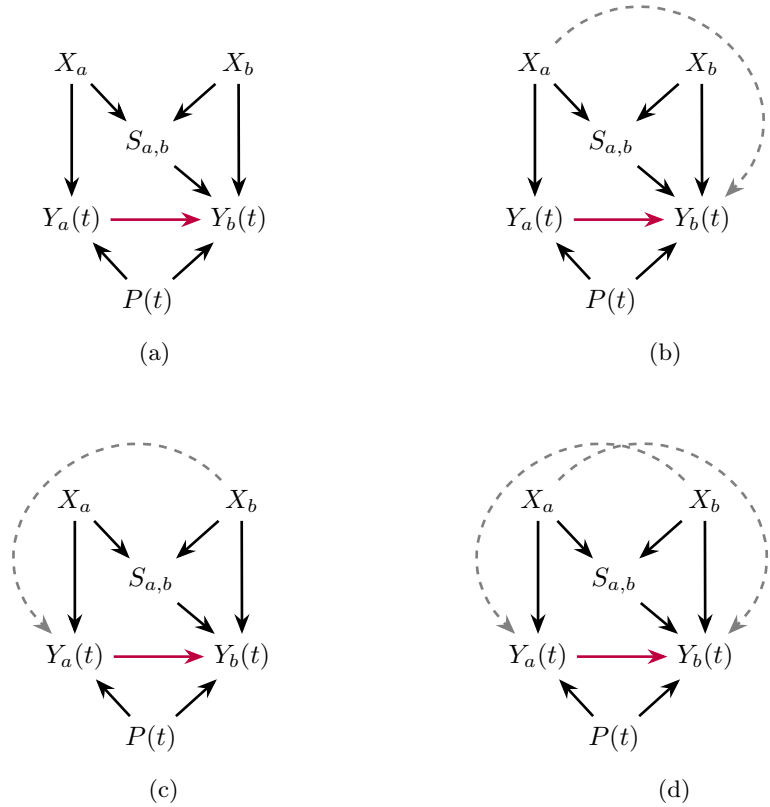

We are interested in measuring the causal path of social influence by observing the outcomes $Y_b(t)$ and $Y_a(t)$ across numerous instances where such outcomes are observed. In particular, we are interested in the chances that the observed outcomes contain identical items, due to the theoretical importance of ``matching'' the social norm and uniformity seeking through behavioral mimicry. We are interested in the causal path of purchasing mimicry, \ie, estimating the causal effect of the treatment ($Y_a(t)$) on the outcome ($Y_b(t)$).

In particular, we observe a person $b$ (focal person), choosing items to purchase at time $t$, $Y_b(t)$. The focal person's choice $Y_b(t)$ is governed by the focal person's eating profile $X_b$. Additionally, we consider common environmental factors in the specific dyad $t$, $P(t)$, that can influence the choices of both observed individuals. Common environmental factors are operationalized as the location, the time of day, popularity, and availability of the item at the shop on the given day.

Furthermore, positioned in front in the queue, before person $b$, there is a frequent peer, person $a$ (partner), choosing items to buy. Similarly, the partner's eating profile $X_a$ impacts their choice $Y_a(t)$. Focal person $b$ can be influenced by person a (partner) in their food choice $Y_a(t)$, corresponding to the causal path of food purchasing mimicry between $Y_a(t)$ and $Y_b(t)$.

The peer's choice can influence the observed person's choice through other biasing paths. In the presence of homophily, the social tie between persons a and b, $S_{a,b}$ is influenced by the traits of each individual $X_a$ and $X_b$, since more similar people tend to be closer friends given homophily, and in turn, influences the observed behavior $Y_b(t)$ through homophilic biasing paths, for closer friendship might make mimicry stronger. Eating profiles composed of habits and preferences are unchanging and independent of individual choices $t$. Social tie strength is a property of the network and is independent of the timing of individual choices $t$. 

In other words, we make the following assumptions:

\begin{assumption}  \label{as:1}
  The traits of partner $X_a$ can influence the observed behavior of the focal person $Y_b(t)$ only through $S_{a,b}$ (we investigate this assumption further in Supplementary Material \Secref{sec:alternativedags} by considering alternative DAGs).
\end{assumption}

\begin{assumption}  \label{as:2}
    We assume that $Y_a(t)$ influences $Y_b(t)$ through the ordering in the queue, while $Y_b(t)$ is not influenced by $Y_a(t)$, i.e., no coordination before purchasing (we investigate this assumption further in Supplementary Material \Secref{sec:coordination}).
\end{assumption}

\begin{assumption}  \label{as:3}
    There are no other unobservable biases (we investigate this assumption further in \Secref{sec:sensitivity} via sensitivity analysis).

\end{assumption}

The causal graph reflecting these assumptions is presented in \Figref{fig:dag1}. The illustrated graph is the standard DAG assumed to identify the causal effect of social influence under the presence of homophily in a pairwise setup when examining the causes behind why a person manifested a behavior at a given time \cite{liotsiou2016social,shalizi2011homophily}. The DAG is equivalent to the causal graph allowing for latent variables to influence both manifest network ties and manifest behaviors when the manifest behaviors are time-independent, \eg, the choices are independent of each other, and there are no other unobservable biases.

According to backdoor criterion \cite{greenland1999causal}, the minimal sufficient adjustment set of variables for estimating the total effect of $Y_a(t)$ on $Y_b(t)$ is \{$X_a$, $P(t)$\}, therefore in our main analyses we match on partner's identity to control for $X_a$ and common environmental factors to control for $P(t)$. In Supplementary Material \Secref{sec:alternativedags} we consider how our estimation framework and the subsequent estimates vary as Assumption \ref{as:1} is violated and additional controls are necessary.

\subsection{Matched estimation framework}\label{matched_framework}

\xhdr{The setup} Given a partner $a$ and a focal person $b$, let $Y_a(t)$ be the partner's choice (set of purchased items within the transaction) and $Y_b(t)$ be the focal person's choice (set of purchased items within the transaction). To estimate the total effect of the partner's purchase on the focal person's purchase ($Y_a(t)$ on $Y_b(t)$), we perform matched estimation. In \Secref{sec:dag}, given the assumed relationship between variables, the sufficient adjustment set of variables is the identity of the partner and the common environmental factors. Common environmental factors are operationalized by measuring the important dimensions of the dietary context: where the food is purchased (shop), when the food is purchased (time), the availability, and the popularity of the food, that date, that time of the day, in that shop as the fraction of all transactions that contained the food item. Availability at the purchase point is approximated by identifying what items were bought at least once for a given date, time of the day, and shop.

\xhdr{Matching} We match dyads in order to find the matched pairs of comparable dyads where in one dyad partner buys the addition $i$ ($i \in Y_a(t)$), whereas in the other partner does not buy the addition $i$ ($i \not\in Y_a(t)$). Within the matched pair of dyads, we ensure that the partner is the same person and that the dyads took place at the same shop and during the same time of the day (breakfast time \vs lunch time \vs afternoon/evening snack time). Additionally, we require that within the matched pair of dyads, the item was available in both dyads and equally popular (up to 10\% caliper), and that both the focal person and the partner purchase the anchor item (meal or a beverage). The size of the popularity caliper was chosen to achieve the balance in covariates, before analyzing the outcomes.

\xhdr{Covariate balance} For all the covariates except food item popularity, an exact match is required. For popularity, we ensured that after matching $SMD < 0.2$ (before matching $SMD=1.23$, after matching $SMD = 0.08$). Groups are considered balanced if all covariates have $SMD<0.2$, a criterion satisfied here \cite{kiciman2018using}.

\xhdr{Outcome analysis} After matching, we analyze 96,986 dyads, matched into 48,493 pairs of dyads. The distributions of dyads across additions are outlined in Table \ref{tab:frequency}. The result is a set of matched pairs of comparable dyads, indistinguishable in the observed attributes, except that in one, the partner buys the additional food item, whereas in the other, the partner does not buy it.

By focusing on different items, we apply our framework to measure the effect of different interventions, in different subpopulations. To quantify the effect of the exposure to the partner's choice, our main analysis compares the purchases of the focal person in the matched pairs of dyads.

Given a food item $i$, $Y_a(t)$ partner's choice (set of purchased items within the transaction) and $Y_b(t)$ focal person's choice (set of purchased items within the transaction), we measure risk difference ($RD_i$) and risk ratio ($RR_i$)~\cite{richardson2017modeling}, calculated based on 2x2 contingency matrix, illustrated in Table \ref{tab:2x2}. The two outcome statistics are defined as:

\begin{equation}\label{eq:rd}
 RD_{i} = p(i \in Y_b(t) | i \in Y_a(t)) - p(i \in Y_b(t) | i \not\in Y_a(t)), 
\end{equation}

and
 
\begin{equation}\label{eq:rr}
RR_{i} = \frac{ p(i \in Y_b(t) | i \in Y_a(t))}{p(i \in Y_b(t) | i \not\in Y_a(t)) }. 
\end{equation}

The risk difference and risk ratio describe the absolute and the relative difference in the observed risk of events between treated and control dyads. For a focal individual, they describe the estimated difference and the relative increase in the probability of purchasing the item. Within the comparable dyads, we resample to obtain the 95\% bootstrapped confidence intervals.

\begin{table}[t!]
\centering
  \caption{\textbf{Contingency table}. The number of dyads in each condition depending on whether or not the partner purchased the item (rows), and whether or not the focal person purchased the item (columns).}
  \label{tab:2x2}
\begin{tabular}{l r c c | c}
    \multicolumn{2}{c}{} & \multicolumn{2}{c}{\textit{Focal purchased}} & \textbf{Total} \\
    %\hline
    \multicolumn{2}{c}{} & \multicolumn{1}{c}{\textbf{No} } & \textbf{Yes} &  \textbf{ dyads} \\
    %\hline
    \textit{Partner} &   \textbf{No} & 40230 &   8263 &  48493 \\
    \textit{purchased} &   \textbf{Yes} & 33332 &  15161 &  48493 \\
    \hline
   \multicolumn{2}{r}{\textbf{Total dyads}}  &73562 &  23424 &  \textbf{96986} \\
  \end{tabular}
\end{table}

\subsection{Amplified asking}\label{aplified_asking}

To make heterogeneous estimates depending on status at the campus (beyond the subpopulation of participants in the sustainability challenge), we rely on the paradigm of amplified asking to, first, build a model that can predict status in the sup-population where the status information is available, and then, second, amplify the entire dataset with the estimated class belonging, by making out-of-sample predictions over the whole population \cite{salganik2019bit}. We train the classifier based on the features that capture temporal patterns typical of personnel and staff. For instance, students make summer and winter breaks, while staff might still be on campus. Similarly, students might make transactions in the later hours.

We use the total number of transactions, the number of years at the campus, and the distribution of transactions across months, weekdays, and hours in the day. The classifier uses a random forest model and achieved, on a 20\% held-out test set, a precision with respect to students of 88.33\% and with respect to personnel of 78.26\%, and a recall with respect to students of 90.60\% and with respect to personnel of 76.60\%. Note that status estimation does not rely on the variables linked with the studied phenomena (purchased items) but merely on the temporal distribution reflecting when the individuals are present on campus.

%\point Calibration analysis demonstrates that although the performance of the classifier is not perfect, the class belonging probability reflects the real prevalence, and can be used for generating estimates based on estimated class belonging probability

% \begin{figure}
%     \centering
%     \begin{minipage}[t]{0.45\textwidth}
%     \centering
%     \includegraphics[width=0.9\textwidth]{fig/fig_1_calibration_plot_students.pdf}
%     \subcaption{}
%     \label{fig:cal_a}
%     \end{minipage}
%     \hfill
%     \begin{minipage}[t]{.45\textwidth}
%     \centering
%     \includegraphics[width = 0.9\textwidth]{fig/fig_1_calibration_plot_staff.pdf}
%     \subcaption{}
%     \label{fig:cal_b}
%     \end{minipage}
% \caption{Status estimation calibration curves. In (a) for student, in (b) for staff prediction. For predicted status probability (on x\hyp axis), the actual fraction of individuals with the status (on \hyp axis). The error bars mark 95\% bootstrapped CI.
% }
% \label{fig:cal}
% \end{figure}

\label{sec:methods}

\section*{Ethical considerations}

Nutrition is a potentially sensitive personal behavior. To protect user privacy, the log data used here was accessed exclusively by EPFL personnel involved in this project, and stored and processed exclusively on EPFL servers. The data was obtained with approval from EPFL's Data Protection Officer and was anonymized before it was made available to the researchers for analysis. Finally, we note that our work was conducted retroactively on data that had been collected passively in order to support campus operations. Thus, our analysis did not influence users in any way.

%\section*{Data availability}

%The individual-level transaction logs cannot be shared publicly. To protect user privacy, the dataset is accessed exclusively by EPFL personnel involved in this project, and stored and processed exclusively on EPFL servers. Upon acceptance, we will make available at a public repository aggregated and anonymized datasets that allow reproducing the main figures and tables.

%\section*{Code availability}

%We make all analysis code publicly available at a public repository.

\bibliography{kristina_phd_thesis_bibliography.bib}

\begin{thebibliography}{91}
\providecommand{\natexlab}[1]{#1}
\providecommand{\url}[1]{\texttt{#1}}
\expandafter\ifx\csname urlstyle\endcsname\relax
  \providecommand{\doi}[1]{doi: #1}\else
  \providecommand{\doi}{doi: \begingroup \urlstyle{rm}\Url}\fi

\bibitem[Ahmed et~al.(2020)Ahmed, Foulkes, Leung, Griffin, Sakhardande,
  Bennett, Dunning, Griffiths, Parker, Kuyken, et~al.]{ahmed2020susceptibility}
S~Ahmed, L~Foulkes, JT~Leung, C~Griffin, A~Sakhardande, M~Bennett, DL~Dunning,
  K~Griffiths, J~Parker, W~Kuyken, et~al.
\newblock Susceptibility to prosocial and antisocial influence in adolescence.
\newblock \emph{Journal of Adolescence}, 84:\penalty0 56--68, 2020.

\bibitem[Althoff et~al.(2022)Althoff, Nilforoshan, Hua, and
  Leskovec]{althoff2022large}
Tim Althoff, Hamed Nilforoshan, Jenna Hua, and Jure Leskovec.
\newblock Large-scale diet tracking data reveal disparate associations between
  food environment and diet.
\newblock \emph{Nature Communications}, 13\penalty0 (1), 2022.

\bibitem[Aral and Nicolaides(2017)]{aral2017exercise}
Sinan Aral and Christos Nicolaides.
\newblock Exercise contagion in a global social network.
\newblock \emph{Nature Communications}, 8\penalty0 (1):\penalty0 1--8, 2017.

\bibitem[Aral et~al.(2009)Aral, Muchnik, and
  Sundararajan]{aral2009distinguishing}
Sinan Aral, Lev Muchnik, and Arun Sundararajan.
\newblock Distinguishing influence-based contagion from homophily-driven
  diffusion in dynamic networks.
\newblock \emph{Proceedings of the National Academy of Sciences (PNAS)},
  106\penalty0 (51), 2009.

\bibitem[Ariely and Levav(2000)]{ariely2000sequential}
Dan Ariely and Jonathan Levav.
\newblock Sequential choice in group settings: Taking the road less traveled
  and less enjoyed.
\newblock \emph{Journal of Consumer Research}, 27\penalty0 (3), 2000.

\bibitem[Ball et~al.(2010)Ball, Jeffery, Abbott, McNaughton, and
  Crawford]{ball2010healthy}
Kylie Ball, Robert~W Jeffery, Gavin Abbott, Sarah~A McNaughton, and David
  Crawford.
\newblock Is healthy behavior contagious: associations of social norms with
  physical activity and healthy eating.
\newblock \emph{International Journal of Behavioral Nutrition and Physical
  Activity}, 7\penalty0 (1), 2010.

\bibitem[Barclay et~al.(2013)Barclay, Edling, and Rydgren]{barclay2013peer}
Kieron~J Barclay, Christofer Edling, and Jens Rydgren.
\newblock Peer clustering of exercise and eating behaviours among young adults
  in sweden: a cross-sectional study of egocentric network data.
\newblock \emph{BMC Public Health}, 13\penalty0 (1), 2013.

\bibitem[Bell et~al.(2019)Bell, Spruijt-Metz, Vega~Yon, Mondol, Alam, Ma, Emi,
  Lach, Stankovic, and De~la Haye]{bell2019sensing}
Brooke~M Bell, Donna Spruijt-Metz, George~G Vega~Yon, Abu~S Mondol, Ridwan
  Alam, Meiyi Ma, Ifat Emi, John Lach, John~A Stankovic, and Kayla De~la Haye.
\newblock Sensing eating mimicry among family members.
\newblock \emph{Translational Behavioral Medicine}, 9\penalty0 (3):\penalty0
  422--430, 2019.

\bibitem[Birch(1980)]{birch1980effects}
Leann~Lipps Birch.
\newblock Effects of peer models' food choices and eating behaviors on
  preschoolers' food preferences.
\newblock \emph{Child Development}, 1980.

\bibitem[Blok et~al.(2013)Blok, van Empelen, Van~Lenthe, Richardus, and
  De~Vlas]{blok2013unhealthy}
DJ~Blok, Pepijn van Empelen, FJ~Van~Lenthe, Jan~Hendrik Richardus, and
  SJ~De~Vlas.
\newblock Unhealthy behaviour is contagious: An invitation to exploit models
  for infectious diseases.
\newblock \emph{Epidemiology \& Infection}, 141\penalty0 (3), 2013.

\bibitem[Bucher et~al.(2016)Bucher, Collins, Rollo, McCaffrey, De~Vlieger,
  Van~der Bend, Truby, and Perez-Cueto]{bucher2016nudging}
Tamara Bucher, Clare Collins, Megan~E Rollo, Tracy~A McCaffrey, Nienke
  De~Vlieger, Daphne Van~der Bend, Helen Truby, and Federico~JA Perez-Cueto.
\newblock Nudging consumers towards healthier choices: a systematic review of
  positional influences on food choice.
\newblock \emph{British Journal of Nutrition}, 115\penalty0 (12), 2016.

\bibitem[Chartrand and Bargh(1999)]{chartrand1999chameleon}
Tanya~L Chartrand and John~A Bargh.
\newblock The chameleon effect: The perception--behavior link and social
  interaction.
\newblock \emph{Journal of personality and social psychology}, 76\penalty0
  (6):\penalty0 893, 1999.

\bibitem[Chartrand and Lakin(2013)]{chartrand2013antecedents}
Tanya~L Chartrand and Jessica~L Lakin.
\newblock The antecedents and consequences of human behavioral mimicry.
\newblock \emph{Annual review of psychology}, 64:\penalty0 285--308, 2013.

\bibitem[Chartrand and Van~Baaren(2009)]{chartrand2009human}
Tanya~L Chartrand and Rick Van~Baaren.
\newblock Human mimicry.
\newblock \emph{Advances in experimental social psychology}, 41:\penalty0
  219--274, 2009.

\bibitem[Christakis and Fowler(2007)]{christakis2007spread}
Nicholas~A Christakis and James~H Fowler.
\newblock The spread of obesity in a large social network over 32 years.
\newblock \emph{New England Journal of Medicine (NEJM)}, 357\penalty0 (4),
  2007.

\bibitem[Christie and Chen(2018)]{christie2018vegetarian}
Chelsea~D Christie and Frances~S Chen.
\newblock Vegetarian or meat? food choice modeling of main dishes occurs
  outside of awareness.
\newblock \emph{Appetite}, 121, 2018.

\bibitem[Collins et~al.(2019)Collins, Thomas, Robinson, Aveyard, Jebb, Herman,
  and Higgs]{collins2019two}
Emily~IM Collins, Jason~M Thomas, Eric Robinson, Paul Aveyard, Susan~A Jebb,
  C~Peter Herman, and Suzanne Higgs.
\newblock Two observational studies examining the effect of a social norm and a
  health message on the purchase of vegetables in student canteen settings.
\newblock \emph{Appetite}, 132, 2019.

\bibitem[Cruwys et~al.(2015)Cruwys, Bevelander, and Hermans]{CRUWYS20153}
Tegan Cruwys, Kirsten~E. Bevelander, and Roel~C.J. Hermans.
\newblock Social modeling of eating: A review of when and why social influence
  affects food intake and choice.
\newblock \emph{Appetite}, 86, 2015.

\bibitem[{de la Haye} et~al.(2010){de la Haye}, Robins, Mohr, and
  Wilson]{DELAHAYE2010161}
Kayla {de la Haye}, Garry Robins, Philip Mohr, and Carlene Wilson.
\newblock Obesity-related behaviors in adolescent friendship networks.
\newblock \emph{Social Networks}, 32\penalty0 (3), 2010.

\bibitem[{de la Haye} et~al.(2011){de la Haye}, Robins, Mohr, and
  Wilson]{DELAHAYE2011719}
Kayla {de la Haye}, Garry Robins, Philip Mohr, and Carlene Wilson.
\newblock How physical activity shapes, and is shaped by, adolescent
  friendships.
\newblock \emph{Social Science and Medicine}, 73\penalty0 (5), 2011.

\bibitem[De~La~Haye et~al.(2011)De~La~Haye, Robins, Mohr, and
  Wilson]{de2011homophily}
Kayla De~La~Haye, Garry Robins, Philip Mohr, and Carlene Wilson.
\newblock Homophily and contagion as explanations for weight similarities among
  adolescent friends.
\newblock \emph{Journal of Adolescent Health}, 49\penalty0 (4):\penalty0
  421--427, 2011.

\bibitem[De~La~Haye et~al.(2013)De~La~Haye, Robins, Mohr, and
  Wilson]{de2013adolescents}
Kayla De~La~Haye, Garry Robins, Philip Mohr, and Carlene Wilson.
\newblock Adolescents' intake of junk food: Processes and mechanisms driving
  consumption similarities among friends.
\newblock \emph{Journal of Research on Adolescence}, 23\penalty0 (3), 2013.

\bibitem[Delaney and McCarthy(2011)]{delaney2011food}
Mary Delaney and Mary McCarthy.
\newblock Food choice and health across the life course: A qualitative study
  examining food choice in older irish adults.
\newblock \emph{Journal of Food Products Marketing}, 17\penalty0 (2-3), 2011.

\bibitem[Deliens et~al.(2016)Deliens, Van~Crombruggen, Verbruggen,
  De~Bourdeaudhuij, Deforche, and Clarys]{deliens2016dietary}
Tom Deliens, Rob Van~Crombruggen, Sofie Verbruggen, Ilse De~Bourdeaudhuij,
  Benedicte Deforche, and Peter Clarys.
\newblock Dietary interventions among university students: A systematic review.
\newblock \emph{Appetite}, 105, 2016.

\bibitem[Eisenberg et~al.(2005)Eisenberg, Neumark-Sztainer, Story, and
  Perry]{EISENBERG20051165}
Marla~E. Eisenberg, Dianne Neumark-Sztainer, Mary Story, and Cheryl Perry.
\newblock The role of social norms and friends' influences on unhealthy
  weight-control behaviors among adolescent girls.
\newblock \emph{Social Science \& Medicine}, 60\penalty0 (6), 2005.

\bibitem[Feunekes et~al.(1998)Feunekes, de~Graaf, Meyboom, and van
  Staveren]{FEUNEKES1998645}
Gerda~I.J. Feunekes, Cees de~Graaf, Saskia Meyboom, and Wija~A. van Staveren.
\newblock Food choice and fat intake of adolescents and adults: Associations of
  intakes within social networks.
\newblock \emph{Preventive Medicine}, 27\penalty0 (5), 1998.

\bibitem[Finnerty et~al.(2010)Finnerty, Reeves, Dabinett, Jeanes, and
  V{\"o}gele]{finnerty2010effects}
Tara Finnerty, Sue Reeves, Jaqueline Dabinett, Yvonne~M Jeanes, and Claus
  V{\"o}gele.
\newblock Effects of peer influence on dietary intake and physical activity in
  schoolchildren.
\newblock \emph{Public Health Nutrition}, 13\penalty0 (3), 2010.

\bibitem[Fjeldsoe et~al.(2011)Fjeldsoe, Neuhaus, Winkler, and
  Eakin]{fjeldsoe2011systematic}
Brianna Fjeldsoe, Maike Neuhaus, Elisabeth Winkler, and Elizabeth Eakin.
\newblock Systematic review of maintenance of behavior change following
  physical activity and dietary interventions.
\newblock \emph{Health Psychology}, 30\penalty0 (1), 2011.

\bibitem[Fletcher et~al.(2011)Fletcher, Bonell, and Sorhaindo]{fletcher2011you}
Adam Fletcher, Chris Bonell, and Annik Sorhaindo.
\newblock You are what your friends eat: systematic review of social network
  analyses of young people's eating behaviours and bodyweight.
\newblock \emph{Journal of Epidemiology \& Community Health}, 65\penalty0 (6),
  2011.

\bibitem[Gakidou et~al.(2017)Gakidou, Afshin, Abajobir, Abate, Abbafati, Abbas,
  Abd-Allah, Abdulle, Abera, Aboyans, et~al.]{gakidou2017global}
Emmanuela Gakidou, Ashkan Afshin, Amanuel~Alemu Abajobir, Kalkidan~Hassen
  Abate, Cristiana Abbafati, Kaja~M Abbas, Foad Abd-Allah, Abdishakur~M
  Abdulle, Semaw~Ferede Abera, Victor Aboyans, et~al.
\newblock Global, regional, and national comparative risk assessment of 84
  behavioural, environmental and occupational, and metabolic risks or clusters
  of risks, 1990--2016: a systematic analysis for the global burden of disease
  study 2016.
\newblock \emph{The Lancet}, 390\penalty0 (10100), 2017.

\bibitem[Garnett et~al.(2019)Garnett, Balmford, Sandbrook, Pilling, and
  Marteau]{garnett2019impact}
Emma~E Garnett, Andrew Balmford, Chris Sandbrook, Mark~A Pilling, and Theresa~M
  Marteau.
\newblock Impact of increasing vegetarian availability on meal selection and
  sales in cafeterias.
\newblock \emph{Proceedings of the National Academy of Sciences (PNAS)},
  116\penalty0 (42), 2019.

\bibitem[Garnett et~al.(2020)Garnett, Marteau, Sandbrook, Pilling, and
  Balmford]{garnett2020order}
Emma~E Garnett, Theresa~M Marteau, Chris Sandbrook, Mark~A Pilling, and Andrew
  Balmford.
\newblock Order of meals at the counter and distance between options affect
  student cafeteria vegetarian sales.
\newblock \emph{Nature Food}, 1\penalty0 (8), 2020.

\bibitem[Gittelsohn et~al.(2012)Gittelsohn, Rowan, and
  Gadhoke]{gittelsohn2012interventions}
Joel Gittelsohn, Megan Rowan, and Preety Gadhoke.
\newblock Interventions in small food stores to change the food environment,
  improve diet, and reduce risk of chronic disease.
\newblock \emph{Preventing Chronic Disease}, 9, 2012.

\bibitem[Gligori{\'c} et~al.(2021)Gligori{\'c}, White, Kiciman, Horvitz,
  Chiolero, and West]{gligoric2021ties}
Kristina Gligori{\'c}, Ryen~W. White, Emre Kiciman, Eric Horvitz, Arnaud
  Chiolero, and Robert West.
\newblock Formation of social ties influences food choice: A campus-wide
  longitudinal study.
\newblock \emph{Proc. of the ACM Conference on Computer Supported Cooperative
  Work and Social Computing (CSCW)}, 2021.

\bibitem[Gligori{\'c} et~al.(2022{\natexlab{a}})Gligori{\'c}, Chiolero,
  K{\i}c{\i}man, White, and West]{gligoric2022population}
Kristina Gligori{\'c}, Arnaud Chiolero, Emre K{\i}c{\i}man, Ryen~W White, and
  Robert West.
\newblock Population-scale dietary interests during the covid-19 pandemic.
\newblock \emph{Nature Communications}, 13\penalty0 (1), 2022{\natexlab{a}}.

\bibitem[Gligori{\'c} et~al.(2022{\natexlab{b}})Gligori{\'c}, Djordjevi{\'c},
  and West]{gligoric2022biased}
Kristina Gligori{\'c}, Irena Djordjevi{\'c}, and Robert West.
\newblock Biased bytes: On the validity of estimating food consumption from
  digital traces.
\newblock \emph{Proc. of the ACM Conference on Computer Supported Cooperative
  Work and Social Computing (CSCW)}, 2022{\natexlab{b}}.

\bibitem[Greenland et~al.(1999)Greenland, Pearl, and
  Robins]{greenland1999causal}
Sander Greenland, Judea Pearl, and James~M Robins.
\newblock Causal diagrams for epidemiologic research.
\newblock \emph{Epidemiology}, pages 37--48, 1999.

\bibitem[Harmon et~al.(2016)Harmon, Forthofer, Bantum, and
  Nigg]{harmon2016perceived}
Brook~E Harmon, Melinda Forthofer, Erin~O Bantum, and Claudio~R Nigg.
\newblock Perceived influence and college students' diet and physical activity
  behaviors: an examination of ego-centric social networks.
\newblock \emph{BMC Public Health}, 16\penalty0 (1), 2016.

\bibitem[Hermans et~al.(2009)Hermans, Larsen, Herman, and
  Engels]{hermans2009effects}
Roel~CJ Hermans, Junilla~K Larsen, C~Peter Herman, and Rutger~CME Engels.
\newblock Effects of social modeling on young women's nutrient-dense food
  intake.
\newblock \emph{Appetite}, 53\penalty0 (1), 2009.

\bibitem[Hermans et~al.(2012)Hermans, Lichtwarck-Aschoff, Bevelander, Herman,
  Larsen, and Engels]{hermans2012mimicry}
Roel~CJ Hermans, Anna Lichtwarck-Aschoff, Kirsten~E Bevelander, C~Peter Herman,
  Junilla~K Larsen, and Rutger~CME Engels.
\newblock Mimicry of food intake: The dynamic interplay between eating
  companions.
\newblock \emph{PloS one}, 7\penalty0 (2):\penalty0 e31027, 2012.

\bibitem[Hetherington et~al.(2006)Hetherington, Anderson, Norton, and
  Newson]{hetherington2006situational}
Marion~M Hetherington, Annie~S Anderson, Geraldine~NM Norton, and Lisa Newson.
\newblock Situational effects on meal intake: A comparison of eating alone and
  eating with others.
\newblock \emph{Physiology \& Behavior}, 88\penalty0 (4-5), 2006.

\bibitem[Higgs and Thomas(2016)]{higgs2016social}
Suzanne Higgs and Jason Thomas.
\newblock Social influences on eating.
\newblock \emph{Current Opinion in Behavioral Sciences}, 9, 2016.

\bibitem[Iacoboni et~al.(1999)Iacoboni, Woods, Brass, Bekkering, Mazziotta, and
  Rizzolatti]{iacoboni1999cortical}
Marco Iacoboni, Roger~P Woods, Marcel Brass, Harold Bekkering, John~C
  Mazziotta, and Giacomo Rizzolatti.
\newblock Cortical mechanisms of human imitation.
\newblock \emph{Science}, 286\penalty0 (5449):\penalty0 2526--2528, 1999.

\bibitem[Jeffery et~al.(1993)Jeffery, Wing, Thorson, Burton, Raether, Harvey,
  and Mullen]{jeffery1993strengthening}
Robert~W Jeffery, Rena~R Wing, Carolyn Thorson, Lisa~R Burton, Cheryl Raether,
  Jean Harvey, and Monica Mullen.
\newblock Strengthening behavioral interventions for weight loss: a randomized
  trial of food provision and monetary incentives.
\newblock \emph{Journal of Consulting and Clinical Psychology}, 61\penalty0
  (6), 1993.

\bibitem[Kakaa et~al.(2018)Kakaa, Bert, Botezatu, Gualano, and
  Siliquini]{eurpub}
O~Kakaa, F~Bert, C~Botezatu, MR~Gualano, and R~Siliquini.
\newblock How we make choices about food? analysis of factors influencing food
  expenditure in northern italy.
\newblock \emph{European Journal of Public Health}, 28, 11 2018.

\bibitem[K{\i}c{\i}man et~al.(2018)K{\i}c{\i}man, Counts, and
  Gasser]{kiciman2018using}
Emre K{\i}c{\i}man, Scott Counts, and Melissa Gasser.
\newblock Using longitudinal social media analysis to understand the effects of
  early college alcohol use.
\newblock \emph{Proc. of the 12th International AAAI Conference on Web and
  Social Media (ICWSM)}, 2018.

\bibitem[Kossinets and Watts(2006)]{kossinets2006empirical}
Gueorgi Kossinets and Duncan~J Watts.
\newblock Empirical analysis of an evolving social network.
\newblock \emph{Science}, 311\penalty0 (5757), 2006.

\bibitem[Krosnick and Alwin(1989)]{krosnick1989aging}
Jon~A Krosnick and Duane~F Alwin.
\newblock Aging and susceptibility to attitude change.
\newblock \emph{Journal of personality and social psychology}, 57\penalty0
  (3):\penalty0 416, 1989.

\bibitem[Levy et~al.(2021)Levy, Pachucki, O’Malley, Porneala, Yaqubi, and
  Thorndike]{levy2021social}
Douglas~E Levy, Mark~C Pachucki, A~James O’Malley, Bianca Porneala, Awesta
  Yaqubi, and Anne~N Thorndike.
\newblock Social connections and the healthfulness of food choices in an
  employee population.
\newblock \emph{Nature Human Behaviour}, 5\penalty0 (10):\penalty0 1349--1357,
  2021.

\bibitem[Liotsiou et~al.(2016)Liotsiou, Moreau, and
  Halford]{liotsiou2016social}
Dimitra Liotsiou, Luc Moreau, and Susan Halford.
\newblock Social influence: From contagion to a richer causal understanding.
\newblock In \emph{International Conference on Social Informatics}, pages
  116--132, 2016.

\bibitem[Madan et~al.(2010)Madan, Moturu, Lazer, and Pentland]{madan2010social}
Anmol Madan, Sai~T Moturu, David Lazer, and Alex~Sandy Pentland.
\newblock Social sensing: Obesity, unhealthy eating and exercise in
  face-to-face networks.
\newblock In \emph{Wireless Health 2010}, 2010.

\bibitem[McCambridge et~al.(2014)McCambridge, Witton, and
  Elbourne]{mccambridge2014systematic}
Jim McCambridge, John Witton, and Diana~R Elbourne.
\newblock Systematic review of the hawthorne effect: New concepts are needed to
  study research participation effects.
\newblock \emph{Journal of clinical epidemiology}, 67\penalty0 (3):\penalty0
  267--277, 2014.

\bibitem[McFerran et~al.(2010)McFerran, Dahl, Fitzsimons, and
  Morales]{doi:10.1086/644611}
Brent McFerran, Darren~W. Dahl, Gavan~J. Fitzsimons, and Andrea~C. Morales.
\newblock I'll have what she's having: Effects of social influence and body
  type on the food choices of others.
\newblock \emph{Journal of Consumer Research}, 36\penalty0 (6), 2010.

\bibitem[Mollen et~al.(2013)Mollen, Rimal, Ruiter, and Kok]{mollen2013healthy}
Saar Mollen, Rajiv~N Rimal, Robert~AC Ruiter, and Gerjo Kok.
\newblock Healthy and unhealthy social norms and food selection. findings from
  a field-experiment.
\newblock \emph{Appetite}, 65, 2013.

\bibitem[Munt et~al.(2017)Munt, Partridge, and
  Allman-Farinelli]{munt2017barriers}
AE~Munt, SR~Partridge, and M~Allman-Farinelli.
\newblock The barriers and enablers of healthy eating among young adults: A
  missing piece of the obesity puzzle: A scoping review.
\newblock \emph{Obesity Reviews}, 18\penalty0 (1), 2017.

\bibitem[Nook and Zaki(2015)]{nook2015social}
Erik~C Nook and Jamil Zaki.
\newblock Social norms shift behavioral and neural responses to foods.
\newblock \emph{Journal of Cognitive Neuroscience}, 27\penalty0 (7), 2015.

\bibitem[Olteanu et~al.(2019)Olteanu, Castillo, Diaz, and
  K{\i}c{\i}man]{olteanu2019social}
Alexandra Olteanu, Carlos Castillo, Fernando Diaz, and Emre K{\i}c{\i}man.
\newblock Social data: Biases, methodological pitfalls, and ethical boundaries.
\newblock \emph{Frontiers in Big Data}, 2, 2019.

\bibitem[Pachucki et~al.(2011)Pachucki, Jacques, and
  Christakis]{pachucki2011social}
Mark~A Pachucki, Paul~F Jacques, and Nicholas~A Christakis.
\newblock Social network concordance in food choice among spouses, friends, and
  siblings.
\newblock \emph{American Journal of Public Health}, 101\penalty0 (11), 2011.

\bibitem[Patrick and Nicklas(2005)]{patrick2005review}
Heather Patrick and Theresa~A Nicklas.
\newblock A review of family and social determinants of children's eating
  patterns and diet quality.
\newblock \emph{Journal of the American College of Nutrition}, 24\penalty0 (2),
  2005.

\bibitem[Pedersen et~al.(2015)Pedersen, Gr{\o}nh{\o}j, and
  Th{\o}gersen]{pedersen2015following}
Susanne Pedersen, Alice Gr{\o}nh{\o}j, and John Th{\o}gersen.
\newblock Following family or friends. social norms in adolescent healthy
  eating.
\newblock \emph{Appetite}, 86, 2015.

\bibitem[Reilly and Kelly(2011)]{reilly2011long}
John~J Reilly and Joanna Kelly.
\newblock Long-term impact of overweight and obesity in childhood and
  adolescence on morbidity and premature mortality in adulthood: systematic
  review.
\newblock \emph{International journal of obesity}, 35\penalty0 (7), 2011.

\bibitem[Reynolds et~al.(2019)Reynolds, Goucher, Quested, Bromley, Gillick,
  Wells, Evans, Koh, Kanyama, Katzeff, et~al.]{reynolds2019consumption}
Christian Reynolds, Liam Goucher, Tom Quested, Sarah Bromley, Sam Gillick,
  Victoria~K Wells, David Evans, Lenny Koh, Annika~Carlsson Kanyama, Cecilia
  Katzeff, et~al.
\newblock Consumption-stage food waste reduction interventions--what works and
  how to design better interventions.
\newblock \emph{Food Policy}, 83, 2019.

\bibitem[Richardson et~al.(2017)Richardson, Robins, and
  Wang]{richardson2017modeling}
Thomas~S Richardson, James~M Robins, and Linbo Wang.
\newblock On modeling and estimation for the relative risk and risk difference.
\newblock \emph{Journal of the American Statistical Association}, 112\penalty0
  (519):\penalty0 1121--1130, 2017.

\bibitem[Robinson and Higgs(2013)]{robinson2013food}
Eric Robinson and Suzanne Higgs.
\newblock Food choices in the presence of ``healthy'' and ``unhealthy'' eating
  partners.
\newblock \emph{British Journal of Nutrition}, 109\penalty0 (4), 2013.

\bibitem[Robinson et~al.(2013)Robinson, Blissett, and
  Higgs]{robinson_blissett_higgs_2013}
Eric Robinson, Jackie Blissett, and Suzanne Higgs.
\newblock Social influences on eating: implications for nutritional
  interventions.
\newblock \emph{Nutrition Research Reviews}, 26\penalty0 (2), 2013.

\bibitem[Robinson et~al.(2014)Robinson, Thomas, Aveyard, and
  Higgs]{ROBINSON2014414}
Eric Robinson, Jason Thomas, Paul Aveyard, and Suzanne Higgs.
\newblock What everyone else is eating: A systematic review and meta-analysis
  of the effect of informational eating norms on eating behavior.
\newblock \emph{Journal of the Academy of Nutrition and Dietetics},
  114\penalty0 (3), 2014.

\bibitem[Rosenbaum(2018)]{rosenbaum2018observation}
Paul Rosenbaum.
\newblock \emph{Observation and experiment}.
\newblock 2018.

\bibitem[Rosenbaum and Silber(2009)]{rosenbaum2009amplification}
Paul Rosenbaum and Jeffrey~H Silber.
\newblock Amplification of sensitivity analysis in matched observational
  studies.
\newblock \emph{Journal of the American Statistical Association}, 104\penalty0
  (488), 2009.

\bibitem[Roy et~al.(2019)Roy, Soo, Conroy, Wall, and
  Swinburn]{roy2019exploring}
Rajshri Roy, Danielle Soo, Denise Conroy, Clare~R Wall, and Boyd Swinburn.
\newblock Exploring university food environment and on-campus food purchasing
  behaviors, preferences, and opinions.
\newblock \emph{Journal of nutrition education and behavior}, 51\penalty0 (7),
  2019.

\bibitem[Salganik(2019)]{salganik2019bit}
Matthew~J Salganik.
\newblock \emph{Bit by bit: Social research in the digital age}.
\newblock 2019.

\bibitem[Salvy et~al.(2008)Salvy, Kieffer, and Epstein]{salvy2008effects}
Sarah-Jeanne Salvy, Elizabeth Kieffer, and Leonard~H Epstein.
\newblock Effects of social context on overweight and normal-weight children's
  food selection.
\newblock \emph{Eating Behaviors}, 9\penalty0 (2), 2008.

\bibitem[Salvy et~al.(2012)Salvy, De~La~Haye, Bowker, and
  Hermans]{salvy2012influence}
Sarah-Jeanne Salvy, Kayla De~La~Haye, Julie~C Bowker, and Roel~CJ Hermans.
\newblock Influence of peers and friends on children's and adolescents' eating
  and activity behaviors.
\newblock \emph{Physiology \& Behavior}, 106\penalty0 (3), 2012.

\bibitem[Sefidgar et~al.(2019)Sefidgar, Seo, Kuehn, Althoff, Browning, Riskin,
  Nurius, Dey, and Mankoff]{sefidgar2019passively}
Yasaman~S Sefidgar, Woosuk Seo, Kevin~S Kuehn, Tim Althoff, Anne Browning, Eve
  Riskin, Paula~S Nurius, Anind~K Dey, and Jennifer Mankoff.
\newblock Passively-sensed behavioral correlates of discrimination events in
  college students.
\newblock \emph{Proc. of the 2019 ACM Conference on Computer Supported
  Cooperative Work and Social Computing (CSCW)}, 2019.

\bibitem[Segovia-Siapco and Sabat{\'e}(2019)]{segovia2019health}
Gina Segovia-Siapco and Joan Sabat{\'e}.
\newblock Health and sustainability outcomes of vegetarian dietary patterns: a
  revisit of the epic-oxford and the adventist health study-2 cohorts.
\newblock \emph{European journal of clinical nutrition}, 72\penalty0 (Suppl
  1):\penalty0 60--70, 2019.

\bibitem[Sen et~al.(2021)Sen, Fl{\"o}ck, Weller, Wei{\ss}, and
  Wagner]{sen2021total}
Indira Sen, Fabian Fl{\"o}ck, Katrin Weller, Bernd Wei{\ss}, and Claudia
  Wagner.
\newblock A total error framework for digital traces of human behavior on
  online platforms.
\newblock \emph{Public Opinion Quarterly}, 2021.

\bibitem[Shalizi and McFowland~III(2016)]{shalizi2016estimating}
Cosma~Rohilla Shalizi and Edward McFowland~III.
\newblock Estimating causal peer influence in homophilous social networks by
  inferring latent locations.
\newblock 2016.

\bibitem[Shalizi and Thomas(2011)]{shalizi2011homophily}
Cosma~Rohilla Shalizi and Andrew~C Thomas.
\newblock Homophily and contagion are generically confounded in observational
  social network studies.
\newblock \emph{Sociological Methods \& Research}, 40\penalty0 (2), 2011.

\bibitem[Sharma et~al.(2010)Sharma, Sivakumaran, and
  Marshall]{sharma2010impulse}
Piyush Sharma, Bharadhwaj Sivakumaran, and Roger Marshall.
\newblock Impulse buying and variety seeking: A trait-correlates perspective.
\newblock \emph{Journal of Business research}, 63\penalty0 (3):\penalty0
  276--283, 2010.

\bibitem[Sharps et~al.(2015)Sharps, Higgs, Blissett, Nouwen, Chechlacz, Allen,
  and Robinson]{sharps2015examining}
Maxine Sharps, Suzanne Higgs, Jackie Blissett, Arie Nouwen, Magdalena
  Chechlacz, Harriet~A Allen, and Eric Robinson.
\newblock Examining evidence for behavioural mimicry of parental eating by
  adolescent females. an observational study.
\newblock \emph{Appetite}, 89:\penalty0 56--61, 2015.

\bibitem[Shepherd(1999)]{shepherd_1999}
Richard Shepherd.
\newblock Social determinants of food choice.
\newblock \emph{Proceedings of the Nutrition Society}, 58\penalty0 (4), 1999.

\bibitem[Singh et~al.(2008)Singh, Mulder, Twisk, Van~Mechelen, and
  Chinapaw]{singh2008tracking}
Amika~S Singh, Chris Mulder, Jos~WR Twisk, Willem Van~Mechelen, and Mai~JM
  Chinapaw.
\newblock Tracking of childhood overweight into adulthood: a systematic review
  of the literature.
\newblock \emph{Obesity reviews}, 9\penalty0 (5), 2008.

\bibitem[Stevenson et~al.(2007)Stevenson, Doherty, Barnett, Muldoon, and
  Trew]{stevenson2007adolescents}
Clifford Stevenson, Glenda Doherty, Julie Barnett, Orla~T Muldoon, and Karen
  Trew.
\newblock Adolescents' views of food and eating: Identifying barriers to
  healthy eating.
\newblock \emph{Journal of Adolescence}, 30\penalty0 (3), 2007.

\bibitem[Sullivan and Huettel(2021)]{sullivan2021healthful}
Nicolette~J Sullivan and Scott~A Huettel.
\newblock Healthful choices depend on the latency and rate of information
  accumulation.
\newblock \emph{Nature Human Behaviour}, 5\penalty0 (12):\penalty0 1698--1706,
  2021.

\bibitem[Swain et~al.(2020)Swain, Kwon, Saket, Morshed, Tran, Patel, Tian,
  Philipose, Cui, Pl\"{o}tz, Choudhury, and Abowd]{swain2020leveraging}
V.~Das Swain, H.~Kwon, B.~Saket, M.~Bin Morshed, K.~Tran, D.~Patel, Y.~Tian,
  J.~Philipose, Y.~Cui, T.~Pl\"{o}tz, M.~De Choudhury, and G.~D. Abowd.
\newblock Leveraging wifi network logs to infer social interactions: A case
  study of academic performance and student behavior, 2020.

\bibitem[Tanner et~al.(2008)Tanner, Ferraro, Chartrand, Bettman, and
  Baaren]{tanner2008chameleons}
Robin~J Tanner, Rosellina Ferraro, Tanya~L Chartrand, James~R Bettman, and
  Rick~Van Baaren.
\newblock Of chameleons and consumption: The impact of mimicry on choice and
  preferences.
\newblock \emph{Journal of Consumer Research}, 34\penalty0 (6):\penalty0
  754--766, 2008.

\bibitem[Thornton et~al.(2013)Thornton, Jeffery, and
  Crawford]{thornton2013barriers}
Lukar~E Thornton, Robert~W Jeffery, and David~A Crawford.
\newblock Barriers to avoiding fast-food consumption in an environment
  supportive of unhealthy eating.
\newblock \emph{Public Health Nutrition}, 16\penalty0 (12), 2013.

\bibitem[Tiefenbeck(2016)]{tiefenbeck2016magnitude}
Verena Tiefenbeck.
\newblock On the magnitude and persistence of the hawthorne effect—evidence
  from four field studies.
\newblock In \emph{4th European Conference on Behaviour and Energy Efficiency,
  Coimbra, Portugal}, pages 8--9, 2016.

\bibitem[Willett et~al.(2019)Willett, Rockstr{\"o}m, Loken, Springmann, Lang,
  Vermeulen, Garnett, Tilman, DeClerck, Wood, et~al.]{willett2019food}
Walter Willett, Johan Rockstr{\"o}m, Brent Loken, Marco Springmann, Tim Lang,
  Sonja Vermeulen, Tara Garnett, David Tilman, Fabrice DeClerck, Amanda Wood,
  et~al.
\newblock Food in the anthropocene: the eat--lancet commission on healthy diets
  from sustainable food systems.
\newblock \emph{The Lancet}, 393\penalty0 (10170), 2019.

\bibitem[Wouters et~al.(2010)Wouters, Larsen, Kremers, Dagnelie, and
  Geenen]{wouters2010peer}
Eveline~J Wouters, Junilla~K Larsen, Stef~P Kremers, Pieter~C Dagnelie, and
  Rinie Geenen.
\newblock Peer influence on snacking behavior in adolescence.
\newblock \emph{Appetite}, 55\penalty0 (1), 2010.

\bibitem[Zhao et~al.(2022)Zhao, Coady, and Bhatia]{zhao2022computational}
Wenjia~Joyce Zhao, Aoife Coady, and Sudeep Bhatia.
\newblock Computational mechanisms for context-based behavioral interventions:
  A large-scale analysis.
\newblock \emph{Proceedings of the National Academy of Sciences}, 119\penalty0
  (15), 2022.

\bibitem[Zipf(2016)]{zipf2016human}
George~Kingsley Zipf.
\newblock \emph{Human behavior and the principle of least effort: An
  introduction to human ecology}.
\newblock Ravenio Books, 2016.

\end{thebibliography}

\section*{Acknowledgments}

We thank Nils Rinaldi, Aurore Nembrini, and Philippe Vollichard for their help in obtaining and anonymizing the transactions. We are also grateful to Jonas Racine and Kiran Garimella for early help with data engineering, and Tim Althoff and Serina Chang for their feedback. This project was funded by the Microsoft Swiss Joint Research Center.

\clearpage

\setcounter{figure}{0}
\setcounter{table}{0}
\setcounter{section}{1}
\setcounter{subsection}{0}
\setcounter{page}{1}
\makeatletter
\renewcommand{\thefigure}{S\arabic{figure}}
\renewcommand{\thetable}{S\arabic{table}}

\section*{Supplementary material}
\subsection{Supplementary information and further robustness checks}

We provide supplementary information and perform analyses that support our main conclusions or provide complementary insights. In Tables \ref{tab:3}, \ref{tab:2}, and \ref{tab:1}, we present co-purchasing matrices that outline the dyad frequency among the subset of the studied
dyads with demographic data available. The tables illustrate a preference for eating with others of the same gender, age, and status.

In \Figref{fig:subpop}, we present the effect estimate among the subpopulation with demographic information available. The estimated risk difference across the matched pairs of dyads is shown separately, depending on the individuals’ status, age, and gender. 

\begin{figure}[b!]
    \centering
    \includegraphics[width=0.7\textwidth]{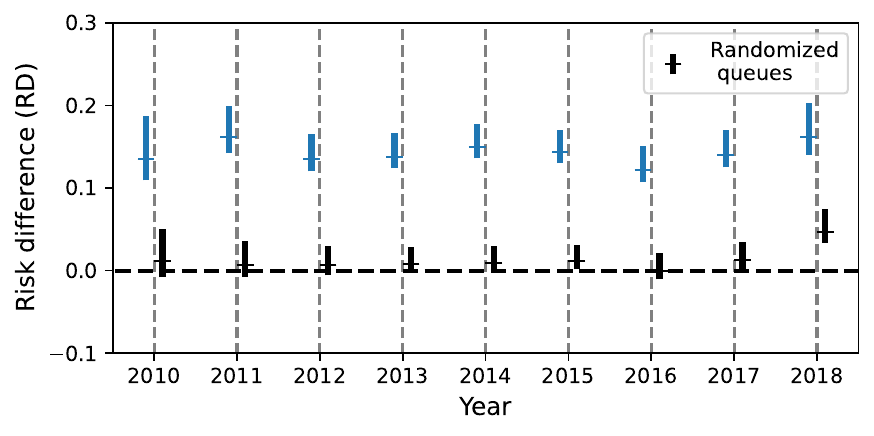}
\caption{\textbf{Effect across the years.} Separately across years (on the x\hyp axis), the estimated risk difference (on the y\hyp axis). The error bars mark 95\% bootstrapped CI. Randomized baseline is presented in black.}
\label{fig:byyear}
\end{figure}

In \Figref{fig:byyear}, we present the estimated risk difference across the nine years spanned by the purchase logs. Over the years, we find no notable trends in the estimated effect. Overall, across 13 food item additions, we measure the highest risk difference in 2011 (16.14\% [14.25\%, 18.05\%]) and the smallest in 2016 (12.21\% [10.73\%, 13.67\%]). In comparison, in the case of the randomized baseline, the highest risk difference is measured in 2018 (4.73\% [3.35\%, 6.13\%]) and the smallest in 2016 (0.00\% [-1.05\%, 1.06\%]). Consistently across years, the partner's influence on the focal person diminishes once the ordering of the queue is randomized.

\begin{figure}[b!]
    \centering
    \includegraphics[width=0.75\textwidth]{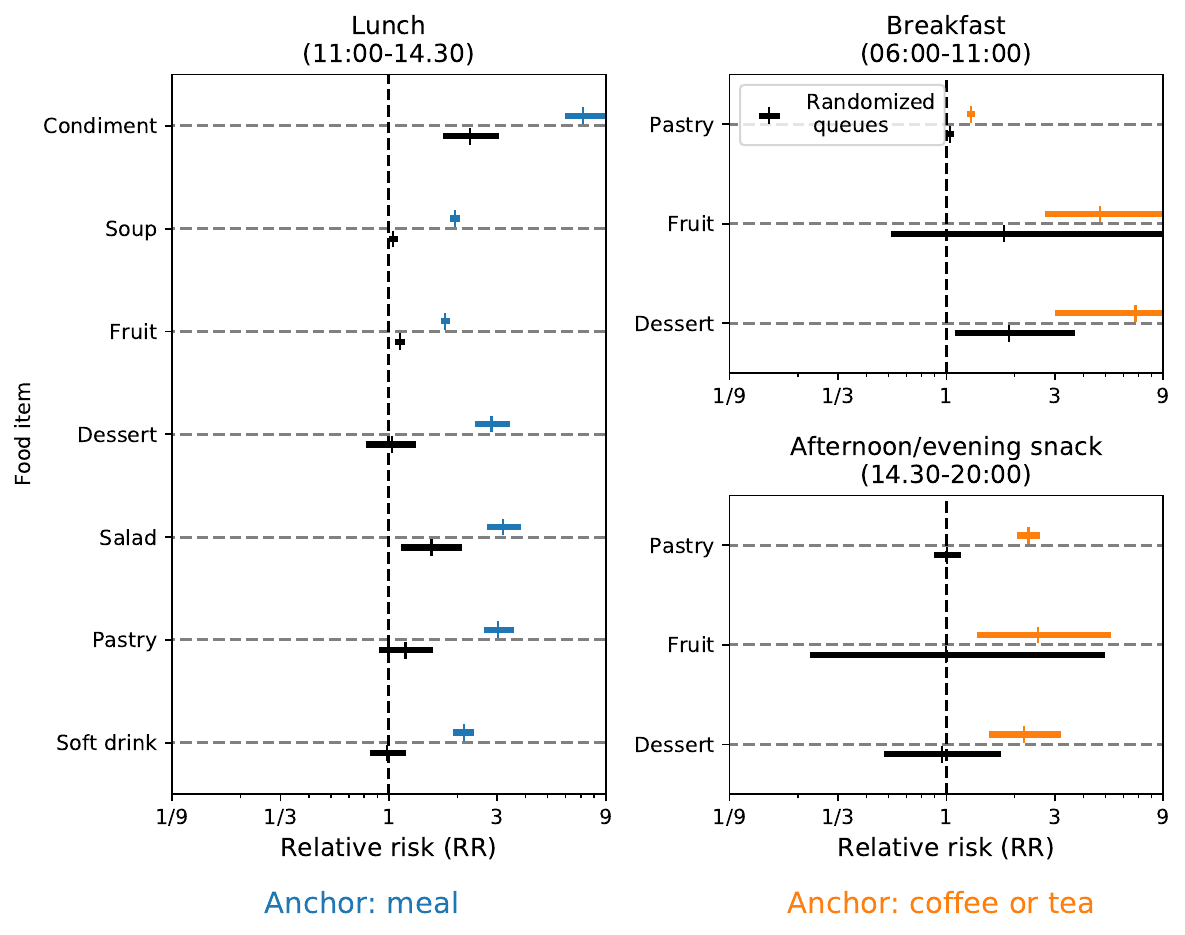}
\caption{\textbf{Risk ratio estimates.} Separately for lunch, breakfast, and afternoon or evening snack, the estimated risk ratio (on the x\hyp axis), for the different food item additions (on the y\hyp axis). The error bars mark 95\% bootstrapped CI. Relative risk estimates are colored (blue for lunch where the anchor is the meal, orange for breakfast and afternoon or evening snack where the anchor is a beverage). Randomized baseline is presented in black. Note the logarithmic x\hyp axis.}
\label{fig:relative}
\end{figure}

Lastly, in \Figref{fig:relative}, we present the relative version of the main findings by measuring the relative risk.

\subsection{Effect estimate under different assumptions}\label{sec:alternativedags}

First, we consider how our estimation framework and the subsequent estimates vary as Assumption \ref{as:1} is violated. The alternative DAGs capture the relaxed assumptions. In \Figref{fig:my_label}, Figures (b), (c), and (d), illustrate the variations of the assumed causal relationships where Assumption \ref{as:1} is violated such that the traits of the individuals can influence the observed purchasing behavior through factors not related to friendship strength $S_{a,b}$. For a set of plausible variations, we derive the minimal sufficient adjustment set of variables according to backdoor criterion \cite{greenland1999causal}. In particular:

\begin{enumerate}
    \item Allowing partner's eating profile to influence the focal person's manifested behaviors through factors not related to friendship strength (\Figref{fig:dag2}), the minimal sufficient adjustment set of variables for estimating the total effect of $Y_a(t)$ on $Y_b(t)$ is \{$X_a$,$P(t)$\}.
    
    \item Allowing focal person's eating profile to influence partners manifested behaviors through factors not related to friendship strength (\Figref{fig:dag3}), the minimal sufficient adjustment set of variables for estimating the total effect of $Y_a(t)$ on $Y_b(t)$ is \{$X_a$,$X_b$,$P(t)$\}.
    
    \item Allowing both eating profiles to influence both manifested behaviors through factors not related to friendship strength (\Figref{fig:dag4}), the minimal sufficient adjustment set of variables for estimating the total effect of $Y_a(t)$ on $Y_b(t)$ is \{$X_a$,$X_b$,$P(t)$\}.
    
\end{enumerate}

\begin{figure}[b!]
    \centering
    \includegraphics[width=0.75\textwidth]{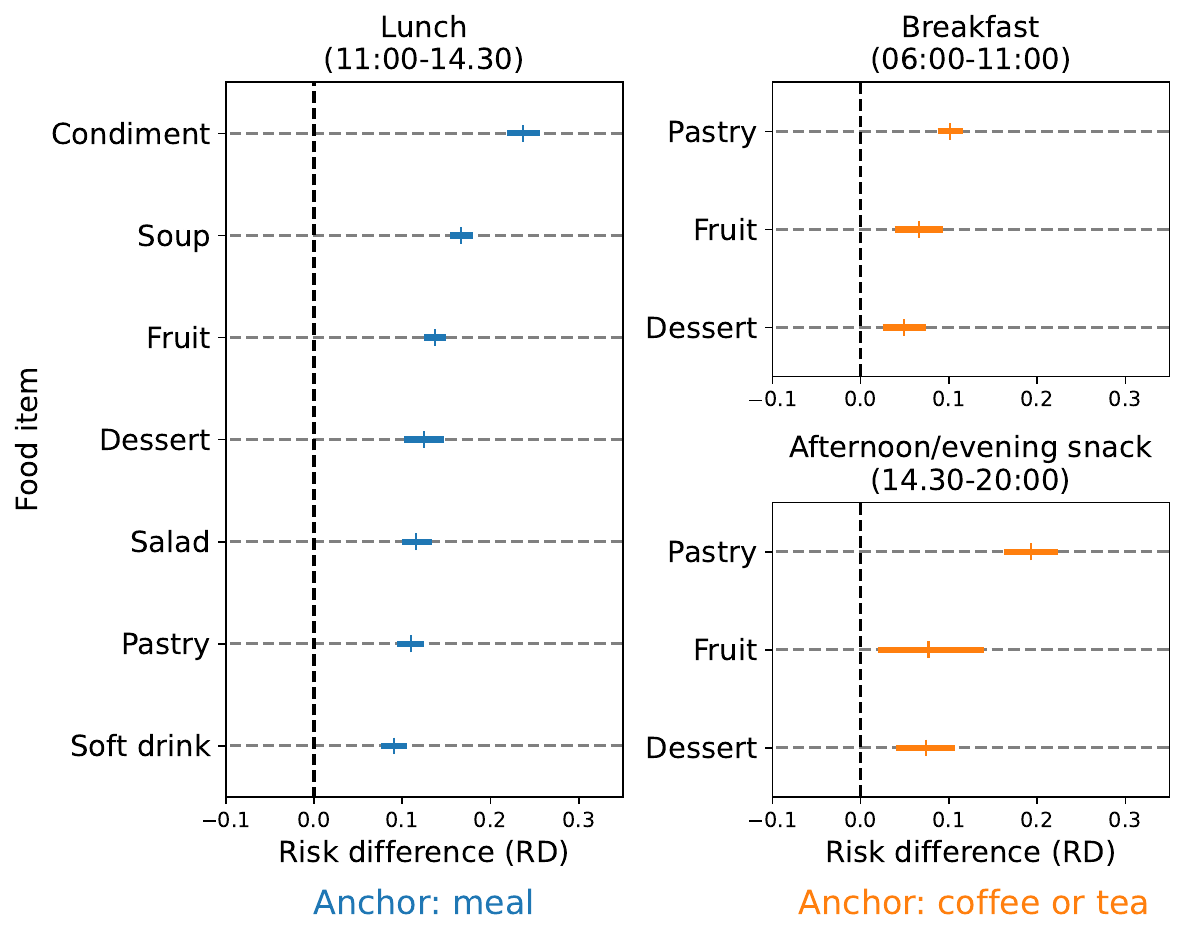}
\caption{\textbf{Effect estimate under different assumptions.} Separately for lunch, breakfast, and afternoon or evening snack, the estimated risk difference (on the x\hyp axis), for the different food item additions (on the y\hyp axis). The error bars mark 95\% bootstrapped CI. Risk difference estimates are colored (blue for lunch where the anchor is the meal, orange for breakfast and afternoon or evening snack where the anchor is a beverage).}
\label{alternative}
\end{figure}

Since the scenario depicted in \Figref{fig:dag2} is already addressed by our main analysis, we investigate how robust estimates are when dyads are additionally matched on focal person identity to control for $X_b$ (necessary in variations depicted in \Figref{fig:dag3} and \Figref{fig:dag4}). When additionally matching on focal person identity, we obtain the overall risk difference of 13.35\% [12.82\%, 13.89\%], risk ratio of 1.76\% [1.72\%, 1.81\%], and qualitatively similar findings as in our main analysis (\cf \Figref{alternative} for risk difference estimate across food items).

\begin{figure}[t!]
    \begin{minipage}[b]{0.32\textwidth}
    \centering
    \includegraphics[width=\textwidth]{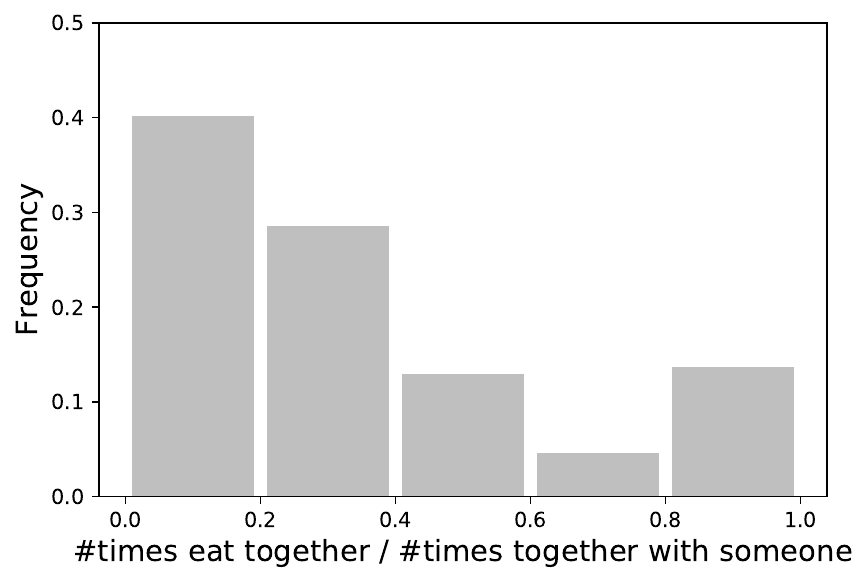}
    \subcaption{}
    \label{fig:normalized_a}
    \end{minipage}
    \hfill
    \begin{minipage}[b]{.32\textwidth}
    \centering
    \includegraphics[width = \textwidth]{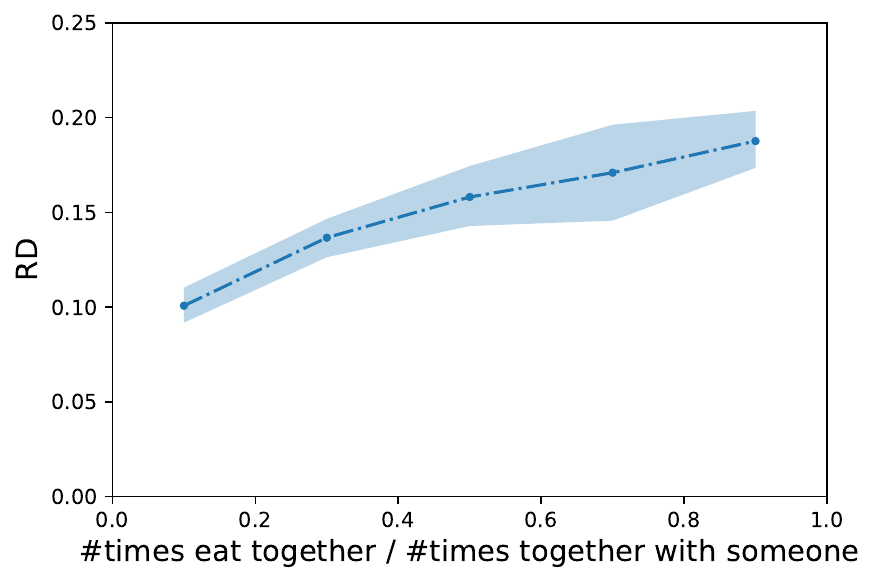}
    \subcaption{}
    \label{fig:normalized_b}
    \end{minipage}
    \hfill
    \begin{minipage}[b]{.32\textwidth}
    \centering
    \includegraphics[width = \textwidth]{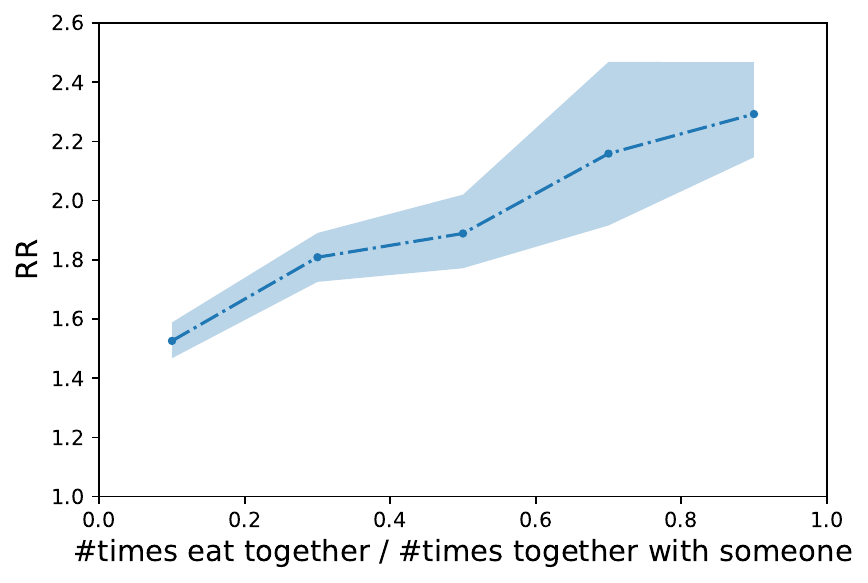}
    \subcaption{}
    \label{fig:normalized_c}
    \end{minipage}
\caption{\textbf{The impact of the tie strength.} In (a), the histogram of the social tie strength between the focal person and the partner. In (b), the risk difference estimate within the subset of matched pairs of dyads (on the y\hyp axis), with the given social tie strengths (on the x\hyp axis). In (b), the risk ratio estimate within the subset of matched pairs of dyads (on the y\hyp axis), with the given social tie strengths (on the x\hyp axis). The shaded areas mark 95\% bootstrapped CI.}
\label{fig:normalized}
\end{figure}

Second, we further investigate the impact of social tie strength. Social tie strength $S_{a,b}$ is operationalized by calculating the fraction of instances when the pair is eating together out of all instances when either one is observed eating with someone. In \Figref{fig:normalized}, we demonstrate that risk difference and risk ratio estimates are the greatest for the highest values of social tie strength. However, the estimates are significant in all strata of social tie strength, and, at minimum, focal persons are estimated to be +10\% more likely to purchase the food item when the partner purchases \vs not.

\subsection{Coordination hypothesis}\label{sec:coordination}

An alternative hypothesis explaining the observed similarities between adjacent persons in the purchasing queue is that the two persons coordinated to go for a meal together and agreed on the food choice before lining up in the purchasing queue. We investigate the presence of such coordination.

There are 226 pairs of persons A and B such that that are at least ten matched pairs of dyads in order A--B and at least ten matched pairs of dyads in order B--A. For each pair, we independently test the coordination null hypothesis that the order A--B or B--A does not matter since similarities come from coordination before making a choice. Under the null hypothesis, people agree on what to eat together before lining up in the queue, so the order of how they go (A--B or B--A) does not make a difference. When the partner purchases an item, the focal person's probability of purchasing is the same in the two orders since the persons pre-agreed, \ie, the purchasing probability does not depend on the order.

Concretely, there is a set of matched pairs of dyads in A--B order and a set of matched pairs of dyads in B--A order. We calculate the purchasing probability of the focal person in the two sets and test the null hypothesis that they are the same. We pool across the 66 pairs by sampling the same number of dyads from each pair (ten), and then perform a two-sided $t$-test. We reject the coordination null hypothesis ($p=3.9\times10^{16}$). 

%When performing a separate test per each pair, we reject ($p<0.05$) the coordination null hypothesis in 66 pairs of persons out of 226 (29\%).

Based on this investigation, we conclude that it is unlikely that pre-purchase coordination can entirely explain the measured effect. Since dyads are matched, differences in ordering likely stem from different mimicry exhibited by person A and person B when they are the focal person \vs\ the partner.

\subsection{Shop layout}\label{layouts}

\begin{figure}[t!]
    \begin{minipage}[b]{0.44\textwidth}
    \centering
    \includegraphics[width=\textwidth]{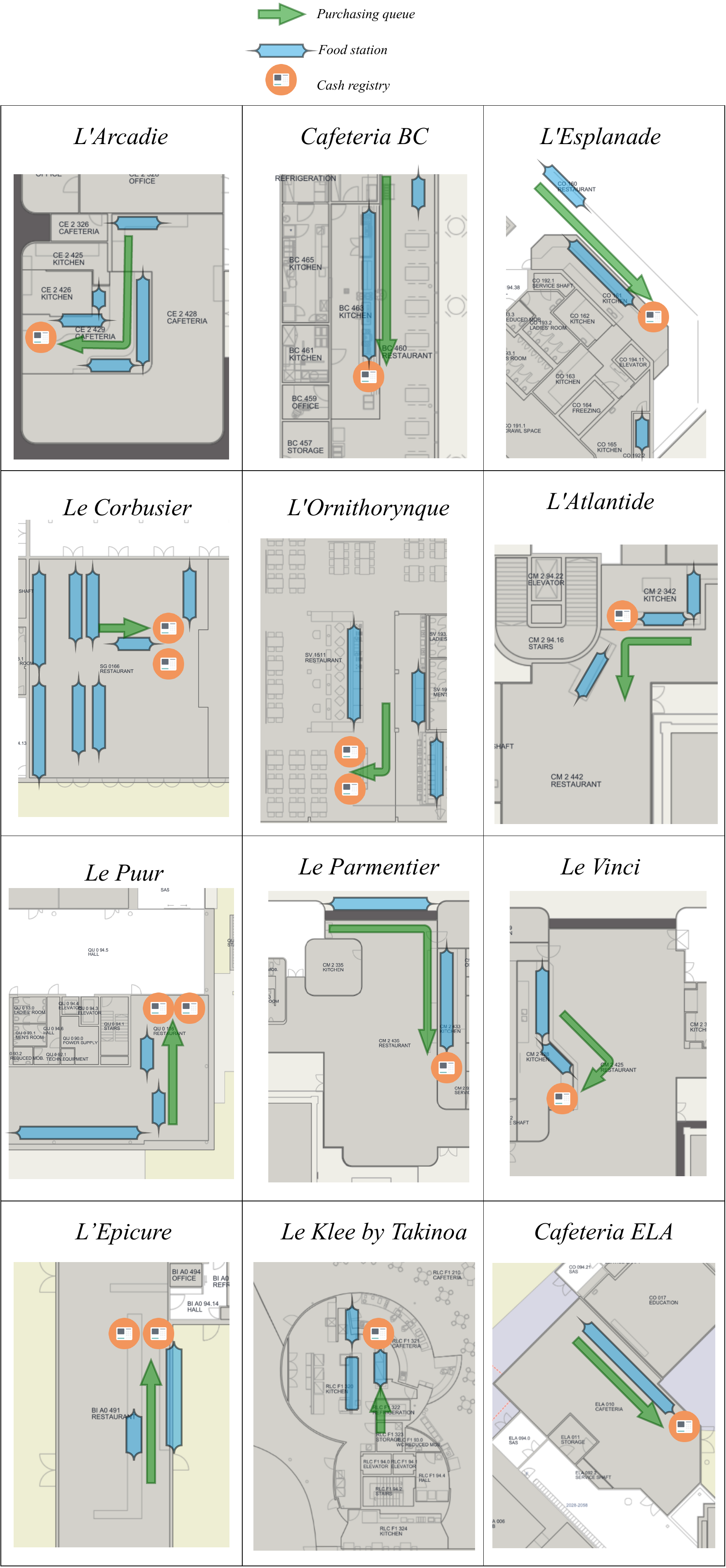}
    \subcaption{}
    \label{fig:layouts_a}
    \end{minipage}
    \hfill
    \begin{minipage}[b]{.55\textwidth}
    \centering
    \includegraphics[width = \textwidth]{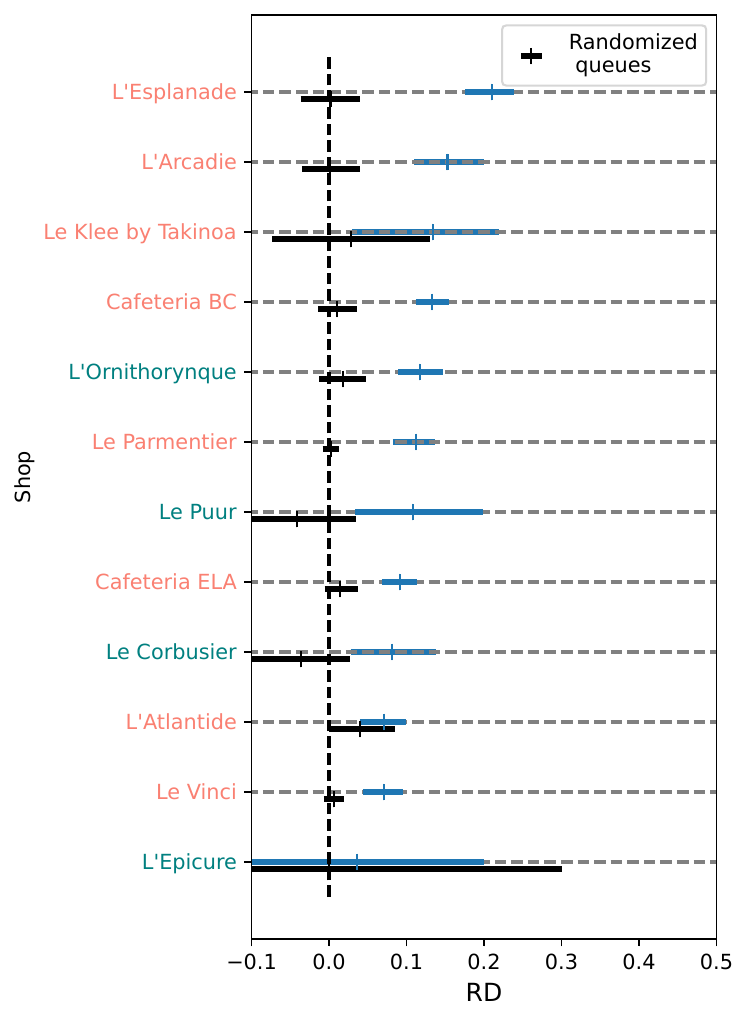}
    \subcaption{}
    \label{fig:layouts_b}
    \end{minipage}
\caption{\textbf{Shop layout.} In (a), visualization of the shop layout based on university map, with purchasing queues (green arrows), food stations (blue rectangles), and cash registries (orange circles) marked. In (b), the estimated risk difference for pastry addition, across the matched pairs of dyads (on the x\hyp axis), across shops (on the y\hyp axis). The error bars mark 95\% bootstrapped CI. Risk difference estimates are presented in blue, the randomized baseline is presented in black. Orange shop names mark shops with a single cash registry, and blue marks shop names with two cash registries.}
\label{fig:layouts}
\end{figure}

In \Figref{fig:layouts}, we visualize the physical layout of the shops and mark the purchasing queues, food stations, and cash registries. Studied food item additions (such as a dessert or a condiment) are placed in font of the cash registry. However, the other food items offered in the shop can be stationed in various layouts (\eg, in a dedicated pasta station, or a salad\hyp meal bar). We expect that shops with multiple cash registry stations display a lower mimicry effect, given that they potentially allow individuals who know each other to split and execute the purchase separately, at different cash registries. 

Since shops vary in food offer, we monitor the estimated effect of mimicking the purchasing of pastry, the most frequent food addition item, available at all the shops (\Tabref{tab:frequency}). We contrast two groups of shops---shops with a single cash registry (8 shops) and shops with two cash registries (4 shops). We measure significantly higher risk difference in shops with a single cash registry, compared to shops with two cash registries (16.77\% [15.93\%, 17.98\%] \vs 13.72\% [13.11\%, 14.33\%]). Furthermore, the four shops with the highest effect estimate are indeed shops with a single cash registry, where food stations tend to be arranged in a straight layout, in front of the cash registry (\Figref{fig:layouts_a}). In summary, we find shops with multiple cash registry stations to have a lower estimated mimicry effect compared to shops with a single cash registry, as predicted.

\subsection{Further controls}\label{further} 
\begin{figure}[b!]
    \centering
    \includegraphics[width=0.75\textwidth]{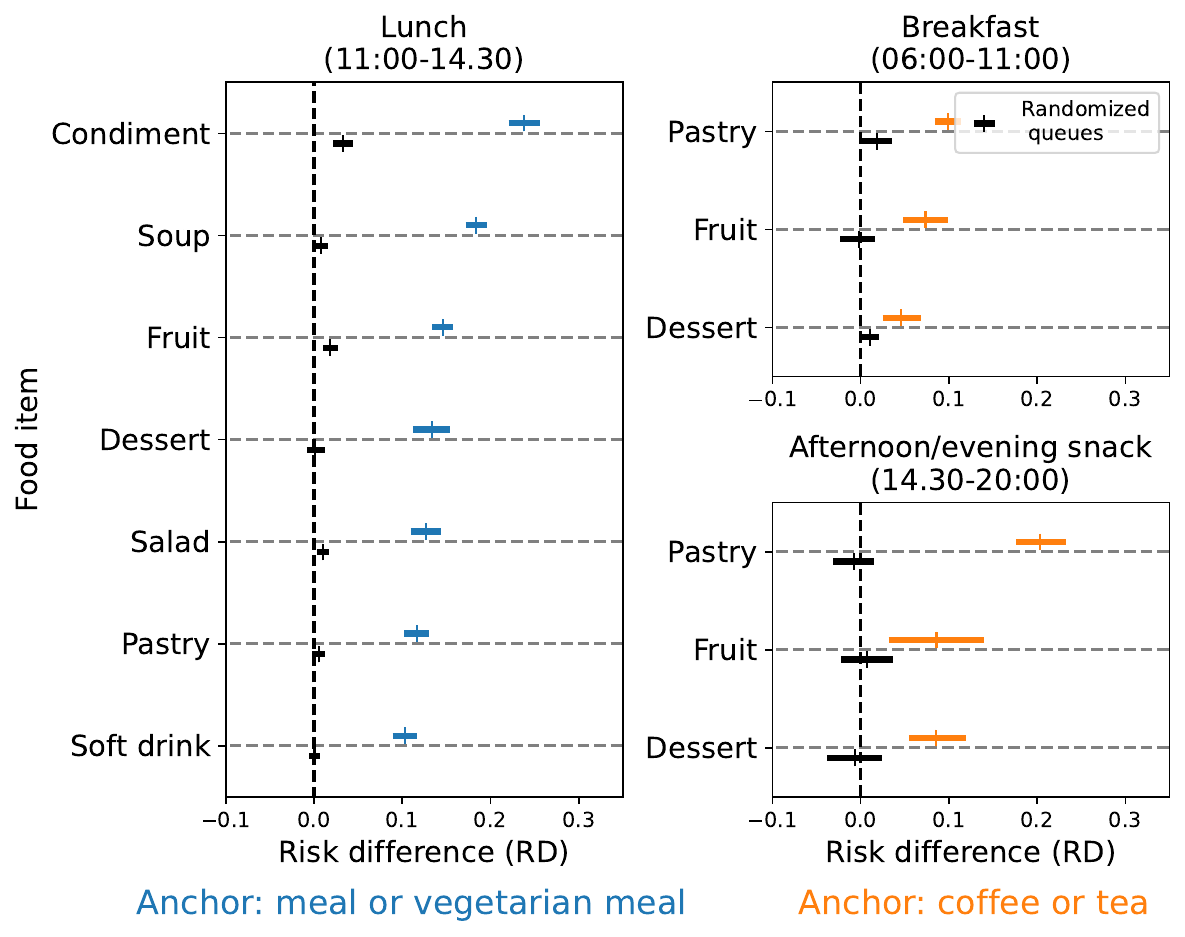}
\caption{\textbf{Absolute effect estimate with additional anchor controls.} Separately for lunch, breakfast, and afternoon or evening snack, the estimated risk difference (on the x\hyp axis), for the different food item additions (on the y\hyp axis). The error bars mark 95\% bootstrapped CI. Risk difference estimates are colored (blue for lunch where the anchor is the meal, orange for breakfast and afternoon or evening snack where the anchor is a beverage).}
\label{additional-diff}
\end{figure}

\begin{figure}[b!]
    \centering
    \includegraphics[width=0.75\textwidth]{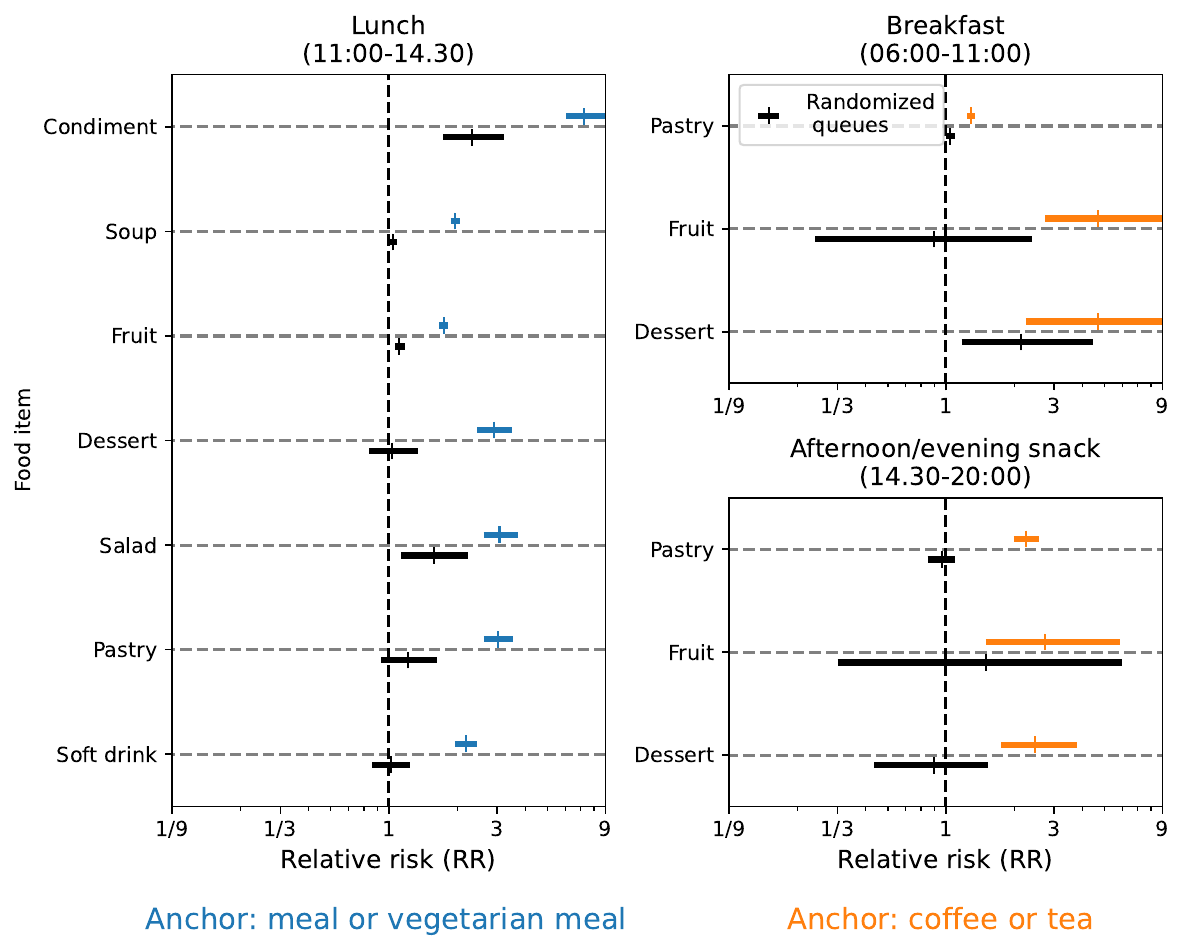}
\caption{\textbf{Relative effect estimate with additional anchor controls.} Separately for lunch, breakfast, and afternoon or evening snack, the estimated risk ratio (on the x\hyp axis), for the different food item additions (on the y\hyp axis). The error bars mark 95\% bootstrapped CI. Relative risk estimates are colored (blue for lunch where the anchor is the meal, orange for breakfast and afternoon or evening snack where the anchor is a beverage).}
\label{additional-ratio}
\end{figure}

We also performed a robustness test additionally requiring that the matched pairs of dyads contain exactly the same anchor (meal \vs vegetarian mean; coffee \vs tea). This separate matching resulted in 94,664 matched pairs of dyads. This led to similar findings as in our main analysis (\cf \Figref{additional-diff} for risk difference and \Figref{additional-ratio} for risk ratio estimates across food items).

\newpage

\begin{table}
    \begin{minipage}{.45\linewidth}
        \caption{ \textbf{Gender co-purchasing matrix.} The condition frequency among the subset of the studied dyads with demographic data available. In rows, the gender of the focal person, in columns, the gender of the partner.}
  \label{tab:3}
 \centering 
\begin{tabular}{l|rr}
\toprule
Partner &  Female &   Male \\
Focal person &         &        \\
\midrule
Female       &   57.00\% &  10.65\% \\
Male         &   12.67\% &  19.68\% \\
\bottomrule
\end{tabular}
    \end{minipage}%
    \hfill
    \begin{minipage}{.45\linewidth}
     \caption{\textbf{Status co-purchasing matrix.} The condition frequency among the subset of the studied dyads with demographic data available. In rows, the status of the focal person, in columns, the status of the partner.}
  \label{tab:2}
  \centering 
\begin{tabular}{lrr}
\toprule
Partner &  Staff &  Student \\
Focal person &        &          \\
\midrule
Staff        &  30.05\% &     9.63\% \\
Student      &   9.83\% &    50.49\% \\
\bottomrule
\end{tabular} 
    \end{minipage} 
\end{table}

\begin{table}
\centering
\begin{minipage}{.6\linewidth}
  \caption{\textbf{Age co-purchasing matrix.} The condition frequency among the subset of the studied dyads with demographic data available. In rows, the age of the focal person, in columns, the age of the partner.}
  \label{tab:1}
\centering 
\begin{tabular}{lrrr}
\toprule
Partner&   $\leq$22 &  23-32 &    >32 \\
Focal person &        &        &        \\
\midrule
$\leq$22      &  25.66\% &   2.34\% &   1.24\% \\
23-32     &   2.67\% &  14.21\% &   9.45\% \\
>32       &   0.02\% &   7.01\% &  37.40\% \\
\bottomrule
\end{tabular}
\end{minipage} 
\end{table}

\begin{figure}[b!]
    \centering
    \begin{minipage}[b]{0.4\textwidth}
    \centering
    \includegraphics[width=\textwidth]{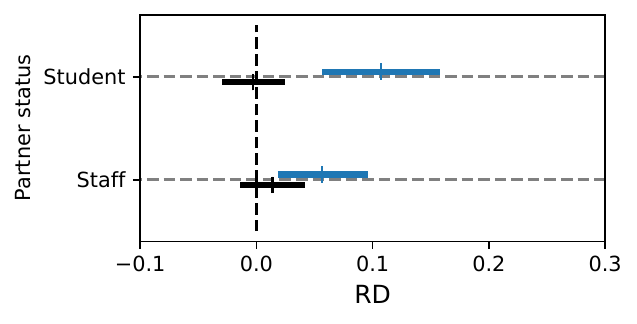}
    \subcaption{}
    \label{fig:subpopa}
    \end{minipage}
    \begin{minipage}[b]{.4\textwidth}
    \centering
    \includegraphics[width = \textwidth]{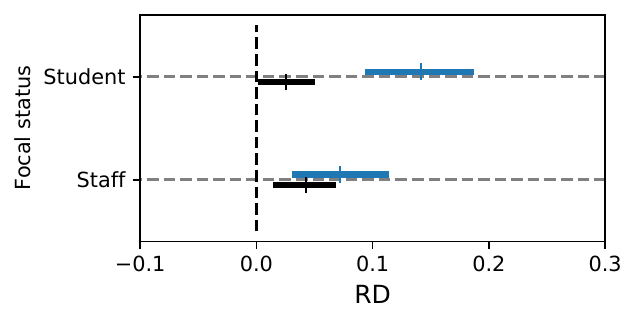}
    \subcaption{}
    \label{fig:subpopb}
    \end{minipage}

    \begin{minipage}[b]{0.4\textwidth}
    \centering
    \includegraphics[width=\textwidth]{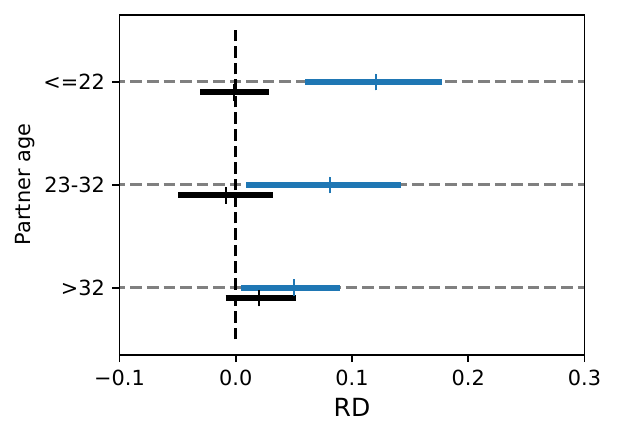}
    \subcaption{}
    \label{fig:subpopc}
    \end{minipage}
    \begin{minipage}[b]{.4\textwidth}
    \centering
    \includegraphics[width = \textwidth]{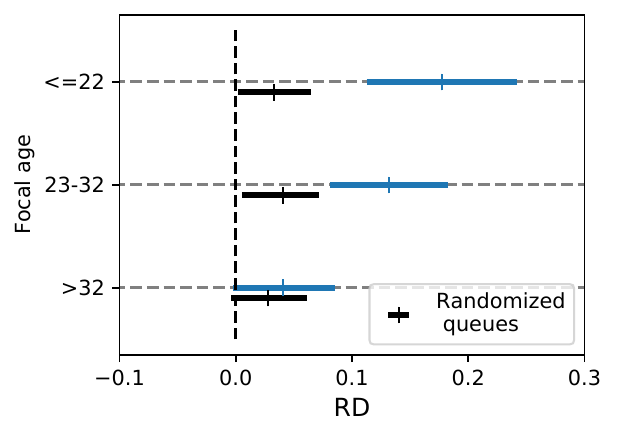}
    \subcaption{}
    \label{fig:subpopd}
    \end{minipage}

    \begin{minipage}[b]{0.4\textwidth}
    \centering
    \includegraphics[width=\textwidth]{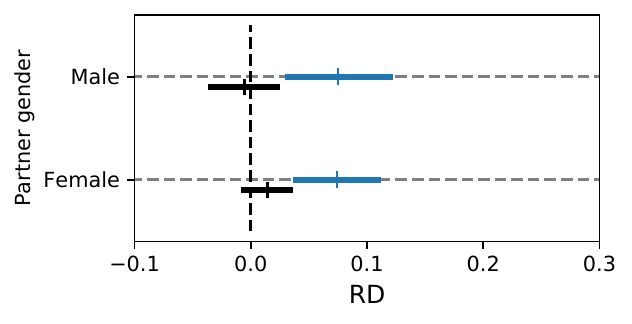}
    \subcaption{}
    \label{fig:subpope}
    \end{minipage}
    \begin{minipage}[b]{.4\textwidth}
    \centering
    \includegraphics[width = \textwidth]{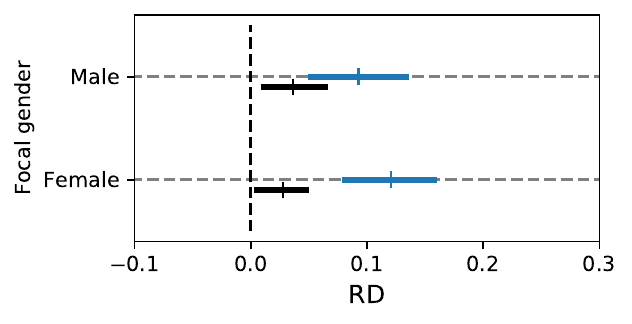}
    \subcaption{}
    \label{fig:subpopf}
    \end{minipage}

\caption{\textbf{Effect by status, age, and gender.} The estimated risk difference across the matched pairs of dyads (on the x\hyp axis), depending on the individuals' status, age, and gender (on the y\hyp axis). The error bars mark 95\% bootstrapped CI. Risk difference estimates are presented in blue, the randomized baseline is presented in black. In (a) for partner's, in (b) for focal person's status, in (c) for partner's, in (d) for focal person's age, in (e) for partner's, and in (f) for focal person's gender.}
\label{fig:subpop}
\end{figure}

%Situations need to happen to break Assumption~\ref{as:1}; you are older, so I leave you the last cake out of politeness

%\point Monitor the same individual, with the same partner; sometimes goes first, sometimes goes second
%\point Intuition: if mimicry: when first, decide autonomously; when second, can be influenced; if coordination: order doesn't matter; when first or second same, because pre-agreed
%\point On average across individuals, we see that when second, different purchasing probability with the same person, compared when first (paired t-test across individuals)
%\point If coordination could completely away explain the effect, we wouldn't expect to see different purchasing probability within the same pair depending on the ordering (ruling out coordination)
%\point Paired t-test ($5.35 \times 10^{-9}$) 0.0473 [0.0461, 0.0486] vs 0.0502 [0.0489, 0.0514]

\label{sec:suppmat}
	
\end{document}